\documentclass[aps,prb,twocolumn,superscriptaddress,longbibliography]{revtex4-2}
\usepackage[colorlinks=true,linkcolor=blue,citecolor=blue]{hyperref}
\usepackage{amsmath,amssymb}
\usepackage[utf8]{inputenc}
\usepackage[T1]{fontenc}
\usepackage{url}
\usepackage{adjustbox}
\usepackage{subcaption}
\usepackage{graphicx}
\usepackage{bm}
\usepackage{bbm}
\usepackage{xcolor}
\usepackage{color}
\usepackage{ulem}
\usepackage{subfloat}
\usepackage{ragged2e}
\usepackage{caption}
\usepackage{braket}
\DeclareCaptionFormat{myformat}{\justifying#1#2#3}
\newsavebox{\measurebox}
\begin{document}

\title{Non-Hermitian Aubry-Andr\'e-Harper model with short- and long-range $p$-wave pairing} 

\author{Shaina Gandhi}
\email{p20200058@pilani.bits-pilani.ac.in}
\affiliation{Department of Physics, Birla Institute of Technology and Science, Pilani 333031, India} 

\author{Jayendra N. Bandyopadhyay}
\email{jnbandyo@gmail.com}
\affiliation{Department of Physics, Birla Institute of Technology and Science, Pilani 333031, India}

\begin{abstract}
We investigate a non-Hermitian Aubry-Andr\'e-Harper model with short-range, as well as with long-range $p$-wave pairing. Here, the non-Hermiticity is introduced through the onsite potential. A comprehensive analysis of several critical aspects of this system is conducted, which includes eigenspectra, topological properties, localization properties, and the transition from real to complex energies. Specifically, we observe the emergence of Majorana zero modes in the case of short-range pairing, whereas the massive Dirac modes emerge in the case of long-range pairing. More importantly, for the case of short-range pairing, we observe two simultaneous phase transitions or double phase transitions: topological and multifractal to localized phase. On the other hand, in case of the long-range pairing, the topological and multifractal to localized transitions do not coincide. However, for both ranges of pairing, we identify a double phase transitions, where delocalized (or metallic) to a critical multifractal state is accompanied by an unconventional shift from real to complex energies. Unlike the short-range pairing case, we observe mobility edges in the long-range pairing case.
\end{abstract}

\maketitle

\section{Introduction}
\label{sec1}

The Aubry-Andr\'e-Harper (AAH) model \cite{AubryAndre, Harper} and the Kitaev chain \cite{A_Yu_Kitaev_2001} are two prominent models which capture significant attention in the community for their distinctive features and contributions in understanding metal-insulator (MI) and topological phase transition. The AAH model describes the dynamics of non-interacting particles in a one-dimensional lattice with a quasi-periodic onsite potential and exhibits MI transition due to its self-dual nature \cite{PhysRevB.55.12971, PhysRevB.91.014108}. 

On the other hand, the Kitaev chain describes the dynamics of 1D spin-less fermions with superconducting $p$-wave pairing. This model has attracted significant interest due to its intriguing topological properties. In particular, the Kitaev chain with short-range pairing exhibits the presence of unpaired Majorana zero modes (MZMs) that are localized at the edges of the chain. This phenomenon also indicates the existence of a topological superconducting phase \cite{RevModPhys.80.1083}. In case of the long-range pairing, the MZMs split into a pair of massive nonlocal edge states known as massive Dirac modes (MDMs) \cite{PhysRevLett.113.156402, Vodola_2016, PhysRevB.94.125121, PhysRevResearch.3.013148}. This massiveness arises because these modes are no longer exactly the zero-energy modes like the MZMs. Instead, due to the non-locality of the pairing between the sites, these modes have a small but finite energy gap. This gap endows these modes with a finite mass and making them massive. The MDMs appear as the mid-gap superconducting states, which indicates that they have energies within the band gap of the bulk excitations of the system. Unlike the MZMs, the MDMs cannot be absorbed into the bulk states of the system. These states are rather topologically nontrivial and propagate along the edges of the system without scattering as long as the system remains in the topologically nontrivial phase. These edge states are a new class of physical quasiparticles that are absent in the standard Kitaev model. The MDMs are the manifestation of the nontrivial topology of the system, which arises due to the presence of the long-range pairing interactions. 

In recent years, there has been considerable interest in studying the AAH model along with the $p$-wave pairing, leading to intriguing findings in the field of quasiperiodic systems \cite{PhysRevResearch.3.013148,PhysRevLett.110.146404, PhysRevLett.110.176403, PhysRevB.93.104504, PhysRevB.94.125408, 10.1093/ptep/ptad043}. Additionally, the exploration of mosaic lattices with $p$-wave pairing  has revealed the emergence of robust MZMs in 1D, which is independent of the strength of spatially modulated potentials \cite{Zeng_2021}. Moreover, the AAH model has been extended to incorporate superconducting pairing in both the incommensurate and commensurate cases. This generalization leads to three distinct topological phases depending on different parameter regimes. These phases include a topologically trivial phase, a Su-Schrieffer-Heeger (SSH)-like topological phase, and a Kitaev-like topological superconducting phase featuring MZMs. These findings hold for the incommensurate \cite{PhysRevB.94.125408}, as well as for the commensurate cases \cite{10.1093/ptep/ptad043}. Besides the short-range pairing, the AAH model with long-range $p$-wave pairing has also been investigated in recent times \cite{PhysRevResearch.3.013148}. This study focuses on the topological properties of the Kitaev chain, considering an AAH onsite potential and algebraically decaying superconducting pairing amplitudes. The exponent of this decay is found to determine a critical strength of pairing, below that the chain remains topologically trivial. In contrast, above this critical value, the system exhibits topological edge modes in the central energy gap. The nature of these edge modes depends on the rate of decay of the pairing term. If the decay is fast enough, the edge modes are identified as MZMs. However, for the slow decay, the edge modes become MDMs \cite{PhysRevB.94.125121}.

More recently, non-Hermitian quantum systems have gained significant interest due to their following unique properties: complex energy spectra \cite{ashida2020non, bergholtz2021exceptional}, exceptional points \cite{wang2021topological, li2020symmetry, yuce2018pt,jin2017schrieffer,zhu2014pt,xu2020fate}, skin effect \cite{PhysRevLett.124.056802,PhysRevLett.124.086801,PhysRevLett.125.126402}, and breaking of conventional bulk-boundary correspondence \cite{yao2018edge,song2019non, lee2016anomalous}. The study of the behavior of the AAH model with non-Hermiticity offers valuable insights into the interplay among quasi-periodicity, topology, and non-Hermiticity \cite{PhysRevA.89.030102, YUCE20142024, PhysRevA.103.L011302, YUCE20151213, PhysRevA.93.062101, longhi2019metal, longhi2019topological, Shu-Chen, PhysRevA.103.033325, PhysRevB.108.014204, PhysRevB.109.L020203, PhysRevResearch.2.033052, PhysRevB.101.020201}. Most of the recent studies of the non-Hermitian quasi-periodic and mosaic potentials in one-dimension focus on investigating localization phenomena, real to complex transition in the spectra, and topological phases in the presence of the $p$-wave superconducting pairing \cite{PhysRevB.105.024514, PhysRevB.103.104203, PhysRevB.103.214202, Cheng_Gao, Xiang_Ping_Jiang}. These investigations aim to understand the interplay between the non-Hermiticity and quasi-periodic structures in the context of $p$-wave superconductivity. However, to the best of our knowledge, the investigation of the non-Hermitian Aubry-Andr\'e-Harper (NH-AAH) model with long-range pairing remains unexplored. In this work, we will carry out a comparative study of the effect of short- and long-range pairing in the NH-AAH model. We also study various aspects for these systems, such as topological properties, localization properties, and real to complex transitions.

This paper is organized as follows: Section \ref{sec2} introduces the model and discusses its general properties. In Sec. \ref{sec3}, we delve into the characteristics of the central gap in the NH-AAH model with $p$-wave pairing, specifically focusing on the comparison between the short- and long-range pairing. Section \ref{sec4} explores the topological properties of the system by analyzing key invariants such as Chern or winding numbers. In Sec. \ref{sec5}, we investigate the localization properties of the system, including the transition from metallic or delocalized to insulating or localized states and an examination of the multifractal behavior of the wave functions. Finally, in Sec. \ref{sec6}, we examine the unconventional real-to-complex transition within the context of both short and long-range pairing. The paper concludes with a summary in Sec. \ref{sec7}.

\section{Model Hamiltonian} 
\label{sec2}

We consider the following fermionic non-Hermitian Hamiltonian on a lattice of length $N$ with nearest-neighbor hopping, long-range $p$-wave superconducting pairing, and onsite AAH modulated chemical potential: 
\begin{equation}
 \begin{split}
 H &= \sum_{j=0}^{N-1}\Bigl[-t(\hat{c}_{j+1}^{\dagger} \hat{ c_j} + \hat{c}_j^\dagger \hat{c}_{j+1})
 - \mu f(j)(2 \hat{c}_j^\dagger \hat{c}_j - 1)\Bigr.\\
 &\Bigl.+ \sum_{l=1}^{N-1} \frac{\Delta}{l^\alpha}(\hat{c}_{j+l}^\dagger \hat{c}_j^\dagger + \hat{c}_j \hat{c}_{j+l})\Bigr].
 \end{split}
 \label{1}
\end{equation}
The non-Hermiticity in the Hamiltonian is introduced by setting the onsite potential $f(j)$ complex as 
\begin{equation}
f(j)= e^{-i(2 \pi \beta j + \theta)}.
\end{equation}
Here, the chemical potential $\mu$ controls the strength of the non-Hermiticity. The parameter $\beta$ is an irrational number, and makes the onsite potential quasiperiodic. The parameter $\theta$ is a phase factor that can be tuned to shift the potential in space, $t$ is the nearest-neighbor hopping amplitude, and $\Delta$ is the superconducting pairing amplitude ($\Delta \in R$). The operators $\hat{c}_j\, (\hat{c}_j^\dagger)$ are the fermionic annihilation (creation) operators at the $j$-th site of the chain. The superconducting pairing follows a power-law decay $l^{-\alpha}$, where $l$ denotes the distance between the sites and the scaling exponent $\alpha \in R^+$. Our model specifically upholds particle-hole symmetry, i.e., $(\mathcal{PC})H(\mathcal{PC})^{-1} = -H$, where parity (spatial reflection) operator is defined as  $\hat{P}^{-1} \hat{c}_j \hat{P} = \hat{c}_{L+1-j}$ and the charge conjugation operator $\mathcal{C}$ is defined as $\mathcal{C}c_j\mathcal{C}^{-1}= ic^\dagger_j$ and $\mathcal{C}i\mathcal{C}^{-1}= -i$. This symmetry leads to the presence of pair of energies ($E$, -$E^*$) in the spectrum. However, in the presence of the $p$-wave pairing term, $\mathcal{PT}$ symmetry is violated \cite{PhysRevB.103.104203,bender1998real, el2018non, bender2002generalized}. 
The Hermitian version of our proposed model, given in Eq. \eqref{1}, is investigated in Ref. \cite{PhysRevResearch.3.013148}. Here, two different forms of $f(j)$ are considered: $f(j) = 1$, constant at all sites; and $f(j) = \cos(2 \pi \beta j + \theta)$, the AAH like quasiperiodic potential. In our model, if we set the parameter $\alpha=0$, then it becomes the NH-AAH model with $p$-wave pairing \cite{PhysRevB.103.104203, PhysRevB.105.024514, PhysRevB.103.214202}. For another case, if we set the chemical potential $\mu=0$ in our model, then it reduces to the standard Kitaev chain with long-range pairing \cite{PhysRevLett.113.156402}. 

A common practice is to set the parameter $\beta = \sigma_G$, where $\sigma_G$ is the inverse of the golden mean, the `most' irrational number and it equals to $(\sqrt{5} - 1)/2$. Interestingly, $\sigma_G = \lim\limits_\infty f_{n-1}/f_n$, a ratio of two consecutive Fibonacci numbers. For numerical calculation, it is convenient to set $\beta = f_{n-1}/f_n$, where $n$ is a finite, but sufficiently large integer. We then set the total number of sites of the system $N = L f_n$, where $L$ is the number of supercells, and each supercell contains $f_n$ sites.
To properly treat the pairing term, the Hamiltonian in Eq. \eqref{1} is diagonalized using Bogoliubov-de Gennes (BdG) basis $\chi_u = (\hat{c}_{f_nu}, \hat{c}_{f_nu}^{\dagger}, \ldots, \hat{c}_{f_nu + (f_n -1)}, \hat{c}_{f_nu + f_n-1}^{\dagger})^{T}$ in real space within a supercell denoted by $u$. Therefore, the Hamiltonian of our model can be written as:
\begin{subequations}
\begin{equation}
 H = \sum_{u, u^\prime=0}^{L-1} \chi_u^\dagger\, H_{u u^\prime}^{\rm BdG}\, \chi_{u^\prime},
\end{equation} 
where 
\begin{equation}
\begin{aligned}
H_{u u^\prime}^{\rm BdG} &= H_{\rm local}\, \delta_{u,u^\prime} + H_{\rm hop}\, \delta_{u+1, u^\prime}\\ &+ \sum_{l=1}^{L-1} H_{\rm pair}^{(l)} \delta_{u+l, u^\prime} + {\rm h.c.}
\end{aligned}
\end{equation}
\label{bdgHam}
\end{subequations} 
and $\delta$'s are Kr\"onecker delta. Here, $H_{\rm local}$ represents the elements of the Hamiltonian involving operators within the supercell, $H_{\rm hop}$ contains the hopping terms between the supercells, and $H_{\rm pair}^{(l)}$ includes the long-range pairing terms. The detailed expressions for these contributions and the Hamiltonian $H^{\rm BdG}$ for the case of a single supercell ($L=1$) incorporating boundary conditions are discussed in Appendix \ref{appendixA}. In this paper, we either use open boundary condition (OBC) to focus on the edge states or anti-periodic boundary condition (APBC) for studying the system's bulk properties. We impose the APBC by setting the condition $\chi_{u+L}= -\chi_u$. By employing the APBC, the cancellation of long-range terms is avoided \cite{PhysRevLett.113.156402}. In our analysis, we focus on a single supercell, i.e., $L=1$, which contains a total of $f_n = 1597$ sites. Therefore, the number of sites in the system is $N = L \times f_n = 1597$. However, for the computation of the winding number, we adopt a method akin to the one utilized in Ref. \cite{PhysRevResearch.3.013148}, wherein $L \rightarrow \infty$ is considered, which allows the study of the system's topological properties in the thermodynamic limit.                          The thermodynamic limit is essential because, in the case of long-range systems, the presence of long-range interactions leads to minimal finite-size effects.
In addition, throughout this paper we set the parameters $t=1.0$ and $\beta=\frac{987}{1597}$. 

\section{Central gap Analysis in the Presence of non-Hermiticity}
\label{sec3} 

First, we examine the characteristics of the central gap in the NH-AAH chain with $p$-wave pairing, particularly comparing the role of short and long-range pairing. For the case of short-range pairing, we set $\alpha = 5.0$, while for the long-range pairing the parameter $\alpha = 0.5$. In this analysis, we investigate the energy spectrum as a function of the chemical potential $\mu$, which determines non-Hermiticity. Here, we set the values of the other parameters as: the pairing strength $\Delta = 0.1t$ and the phase $\theta = 0$. 

\begin{figure*}[t]
    \centering
\begin{tabular}{cc}
    \vspace{-0.3cm}
    \hspace{-0.5cm}
    \includegraphics[width=0.495\textwidth]{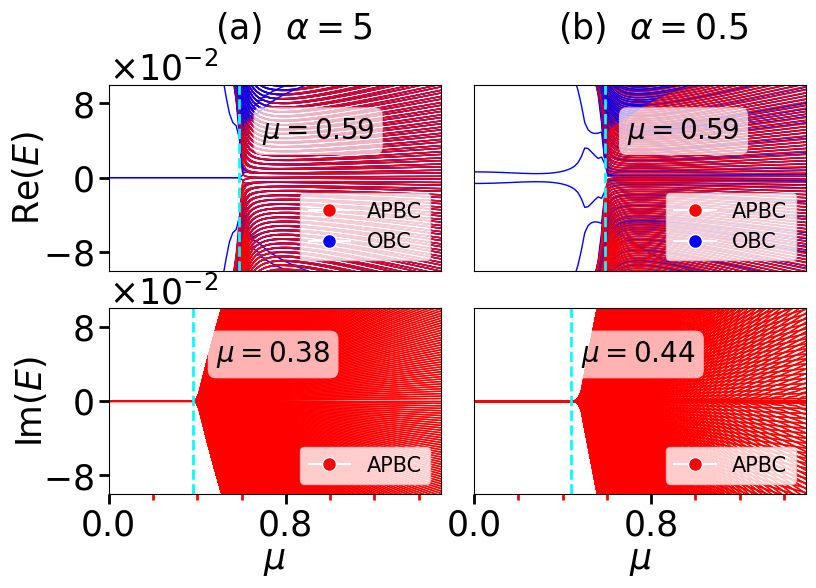} &
    \hspace{-0.2cm}
    \includegraphics[width=0.4\textwidth]{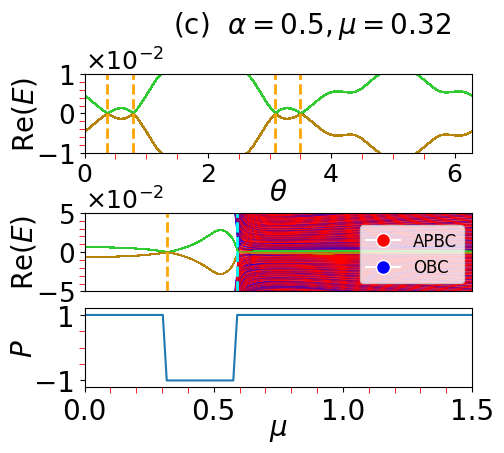}  
\end{tabular}
\caption{(a, b) Real and imaginary parts of eigenvalues for the system with OBC (blue) and APBC (red) are depicted as a function of the non-Hermitian parameter $\mu$ for a system size $N = 1597$. The other system parameters are set at $\Delta = 0.1$, and $\theta = 0$. (a) The MZMs are identified within the central gap region $0 \leq \mu \lesssim 0.59$ by examining the real part of the eigenvalues. For $0 \leq \mu \lesssim 0.38$, the imaginary part of the eigenspectra is fixed at {\it zero}, which indicates that the spectra is real in this range. (b) The presence of MDMs at the central region $0 \leq \mu \lesssim 0.59$ is evident after analyzing the real part of the eigenspectra. However, the imaginary part of the eigenspectra reveals that the eigenvalues become real for $0 \leq \mu \lesssim 0.44$. (c) The real part of the eigenvalues is shown as a function of the phase factor $\theta$ for $\mu = 0.32$ and $\Delta = 0.5$. The green color represents the edge mode with positive energy, while the golden color represents the edge mode with negative energy. Orange dashed lines mark crossings of these edge modes at $\theta =  0.37$, $0.79$, $3.09$, and $3.50$, respectively. In the middle subfigure, real parts of the energy eigenvalues are presented as a function of the parameter $\mu$ for $\theta = 0.79$. A crossing is observed at $\mu = 0.32$, indicated by the orange-colored dashed line. In the bottom subfigure, the ground state fermion parity is presented. At the specific points, where the energy spectrum shows a crossing, notably at $\mu = 0.32$, and at the gap closing point $\mu = 0.59$, the fermion parity of the ground state undergoes a switch.}
  \label{Eigenspectra}
\end{figure*}

Figure \ref{Eigenspectra} depicts the real and imaginary part of the energy spectrum of the system for a modulated AAH potential with OBC (blue) and APBC (red) for the short-range ($\alpha = 5.0$), as well as for the long-range pairing ($\alpha = 0.5$). In the top panel of Fig. \ref{Eigenspectra}(a), for the case of short-range pairing, the emergence of MZMs is evident within the central energy gap for the parameter range $0 \leq \mu \lesssim 0.59$. Examining the imaginary part of the eigenvalues, given in the bottom panel of  Fig. \ref{Eigenspectra}(a), we find that the eigenspectra exhibits real behavior for $0 \leq \mu \lesssim 0.38$. For the long-range system, the top panel of Fig. \ref{Eigenspectra}(b) displays the appearance of MDMs in the central energy gap for $0 \leq \mu \lesssim 0.59$.  
In Appendix \ref{appendixB}, we demonstrate that, for the pairing strength $\Delta = 0.5$, the MZMs and MDMs do not appear within the same parameter regime. Additionally, while examining the imaginary part of the eigenvalues in the bottom panel of Fig. \ref{Eigenspectra}(b), we find that the eigenvalues become real for $0 \leq \mu \lesssim 0.44$. 

We also present the real part of the energy as a function of the phase angle $\theta$ for the long-range pairing case in the top panel of Fig. \ref{Eigenspectra}(c). For this analysis, we have fixed the parameters at $\Delta = 0.1$, $\alpha = 0.5$, and $\mu = 0.32$. We specifically set this value of the parameter $\mu$ because we have observe in Fig. \ref{Eigenspectra}(b) that the gap between the two MDMs is the smallest at this particular value. We also note that the MDMs exhibit crossings at $\theta =  0.37,\, 0.79,\, 3.09$, and $3.50$. Here, the lowest positive eigenvalue is shown in green, whereas the highest negative value is shown in golden color. The vertical orange lines indicate the values of $\theta$ at which the crossings occur. These crossing points establish the adiabatic connection between the MZMs and the MDMs \cite{PhysRevResearch.3.013148}. The presence or absence of this connection depends on the phase value of the AAH potential \cite{PhysRevB.94.125121}. In addition, we have also presented the real part of the eigenvalues for the same parameters as in Fig. \ref{Eigenspectra}(b), but for a phase angle $\theta = 0.79$. Here, we observe a crossing at $\mu = 0.32$. Furthermore, following Ref. \cite{PhysRevB.94.115166}, we have plotted the ground state fermion parity's behavior to distinguish the genuine crossings from the avoided ones. The ground state fermion parity is defined as:
\begin{equation}
P = {\rm sgn}[{\rm Pf}(H_M)].
\label{Pf}
\end{equation}
Here, ${\rm Pf}$ denotes the Pfaffian \cite{Wimmer}, while $H_M$ denotes the Hamiltonian in the Majorana basis with OBC. In the Majorana basis, quantum states are expressed in terms of the Majorana operators. The fermionic operators $c_j$ and $c_j^\dagger$ can be expressed in terms of the Majorana operators $a_{2j}$ and $a_{2j-1}$ as:
\[
c_j = \frac{a_{2j} + a_{2j-1}}{2i} \quad \text{and} \quad c_j^\dagger = -\frac{a_{2j} - a_{2j-1}}{2i},
\]
where $j = 1, 2, ..., N$. These expressions satisfy the following relations: $a_j^\dagger = a_j$ and $a_l a_m + a_m a_l = 2 \delta_{lm}$, where $l, m = 1, 2, ..., 2N$ \cite{A_Yu_Kitaev_2001}. The representation of the Hamiltonian in the Majorana basis is crucial, because it transforms the Hamiltonian as a skew-symmetric matrix, which is essential prerequisite for the computation of Pfaffian. The bottom panel of Fig. \ref{Eigenspectra}(c) shows that the fermion parity changes sign at the crossing point $\mu = 0.32$ and also at the gap closing point $\mu = 0.59$. This behavior confirms the exchange of the two orthogonal modes at the both points, and consequently affirms the genuine nature of the level crossings rather than the avoided crossings.

\section{Topological transition} 
\label{sec4}

Here we adopt the methodology employed in Ref. \cite{PhysRevResearch.3.013148} to characterize the non-Hermitian topological superconducting phases in the 1D quasiperiodic lattice with short- and long-range $p$-wave pairing. We calculate the system's winding number in the thermodynamic limit as $L \rightarrow \infty$. This approach provides a reliable and consistent way to assess the topological properties of the system across the phase transition. Here, we analyze the system properties in the momentum space. We perform a Fourier transformation by expressing the Hamiltonian of the system in the {\it BdG momentum basis}, given by $\chi_k = (c_{k,0}, c_{-k,0}^{\dagger}, \ldots, c_{k ,f_n -1}, c_{-k + f_n-1}^{\dagger})^{T}$. The Hamiltonian at $L \rightarrow \infty$ limit is denoted by $H_{\text{inf}}$. The explicit form of $H_{\text{inf}}$, as well as the details of the Fourier transformation, are provided in Appendix \ref{appendixC}.

\begin{figure}[b]
\centering
\includegraphics[width=0.48\textwidth]{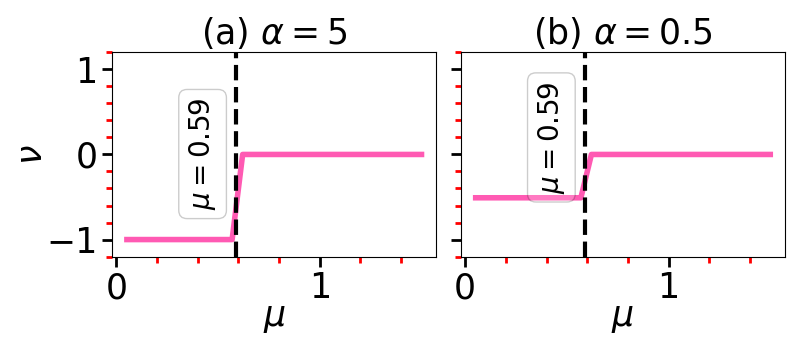} 
\caption{Variation of winding number $\nu$ is presented as a function of the the non-Hermitian parameter $\mu$. Subfigures (a) and (b) are respectively plotted for the short- and the long-range pairing cases. The black dashed line marks the topological phase transition point. In (a), the winding number $\nu$ undergoes a transition from $-1$ to $0$ around $\mu = 0.59$. On the other hand, (b) illustrates a transition in the winding number $\nu$ from $-0.5$ to $0$ again around the same $\mu = 0.59$.}
\label{winding}
\end{figure}

One can also represent the Hamiltonian $H_{\text{inf}}$ in a block-off-diagonal form by rearranging the BdG momentum basis as $\chi_k^\prime = (c_{k,0},\ldots c_{k,f_n-1}, \ldots, c_{-k ,0}^{\dagger},\ldots c_{-k + f_n-1}^{\dagger})^{T}$. In this $\chi_k^\prime$ basis, the Hamiltonian $H_{\text{inf}}^\prime$ becomes:
\begin{equation}
H_{\text{inf}}^\prime = \begin{pmatrix} 0 & h \\ h^{\dagger} & 0 \end{pmatrix}
\end{equation}
We compute the winding number using the relation given in \cite{PhysRevX.9.041015}:
\begin{equation}
\nu = \int_0^{2\pi} \frac{1}{2\pi i} \mathrm{Tr}\left[h^{-1} \partial_{k} h\right].
\end{equation}
Here, the winding number is calculated for both short- and long-range pairing cases as a function of the non-Hermitian parameter $\mu$. The results are presented in Fig. \ref{winding}. From a rigorous numerical study of the system, we identify topological phase transition at $\mu = 0.59$ for both short- and the long-range pairing. As depicted in Fig. \ref{Eigenspectra}(a, b), the presence of MZMs and MDMs is evident in the regime $\mu \leq 0.59$ for the both ranges of pairings. Thus, the MZMs have the winding number $\nu = -1$, while the winding number of the MDMs is $\nu = -\frac{1}{2}$ \cite{PhysRevB.94.125121}. This result shows that the topological phase transition and the superconducting phase transition occur simultaneously for both the range of pairing. 

\section{Localization and (multi)fractal properties} 
\label{sec5}

\begin{figure*}[t]
    \centering
    \begin{tabular}{ccc}
        \includegraphics[width=0.25\linewidth]{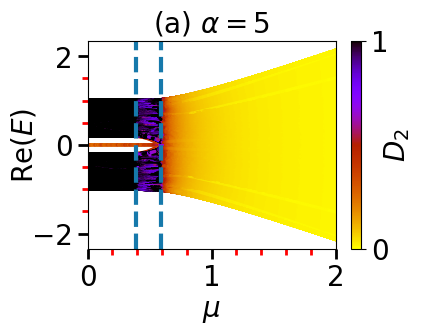} & 
        \includegraphics[width=0.44\linewidth]{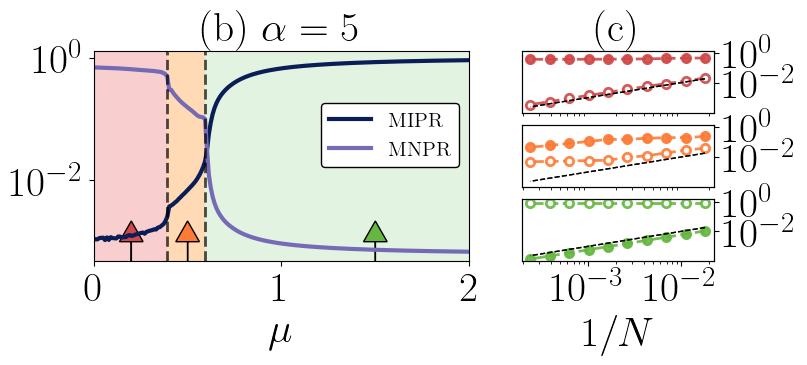} & 
        \includegraphics[width=0.3\linewidth]{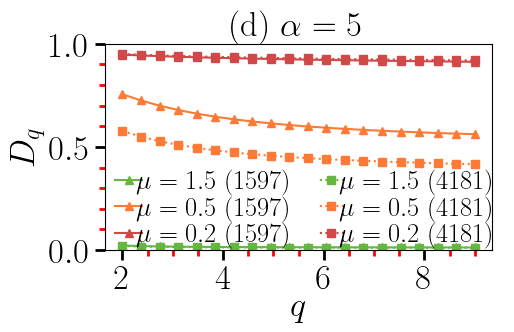} \\
        
        \includegraphics[width=0.25\linewidth]{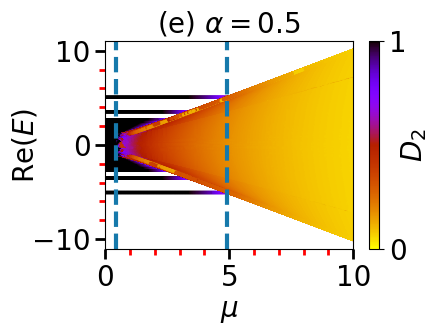} & 
        \includegraphics[width=0.44\linewidth]{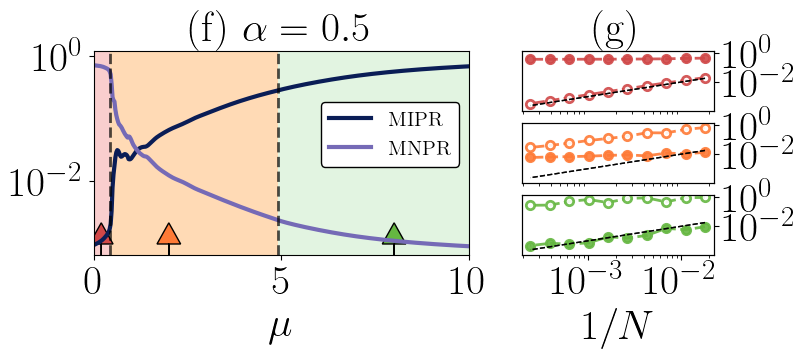} & 
        \includegraphics[width=0.3\linewidth]{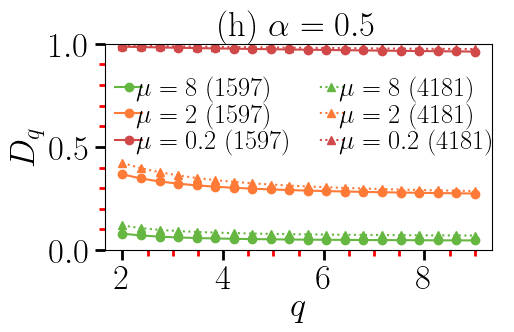} \\
    \end{tabular}
    \caption{The figure's top row displays the short-range pairing case results with $\alpha = 5.0$. The bottom row presents the results for the long-range pairing case with $\alpha = 0.5$. Subfigures (a) and (e) illustrate the behavior of the fractal dimension $D_2$ of the energy eigenstates with the pairing strength $\Delta = 0.1$. The blue-colored dashed lines separate delocalized, (multi)fractal, and localized regions. In subfigures (b) and (f), the logarithm of MIPR and MNPR is depicted. In (b), the black dashed lines at $\mu = 0.38$ and $\mu = 0.59$ represent the boundaries of different phases, while in (f), the phase boundaries are at $\mu = 0.44$ and $\mu = 4.9$. In these subfigures, we mark the following values of $\mu$ by arrows, where we have shown the finite-size scaling of the MIPR and MNPR. For the short-range pairing, arrows denote $\mu = 0.2$ (red), $\mu = 0.5$ (orange), and $\mu = 1.5$ (green); whereas for the long-range pairing, arrows correspond to $\mu = 0.2$ (red), $\mu = 2.0$ (orange), and $\mu = 8.0$ (green). Subfigures (c) and (g) demonstrate the finite-size scaling behaviors of the MIPR (open circles) and MNPR (filled circles) for the system sizes $N = 34$ to $4181$. The colors of these plots are indicating the corresponding values of the parameter $\mu$, which are marked by the same colored arrows in subfigures (b) and (c). Finally, subfigures (d) and (h) show the generalized fractal dimensions for the same $\mu$ values. Here, for the multifractal cases, $D_q$ has fractional values and is also dependent on $q$.}
    \label{localization}
\end{figure*}

We apply the traditional box counting procedure to calculate the (multi)fractal dimension of the eigenstates to study the localization property of the system \cite{PhysRevLett.67.607}. The system which is considered here has $N$ sites, then the Hamiltonian will be a $N \times N$ matrix, and consequently its eigenstates will be $N$-dimensional. We calculate the fractal dimension of the eigenstates by the following way: we divide the components of the eigenstates into $N_d$ boxes, and hence each box will have $d=N/N_d$ number of components. If the $i$-th eigenstate of the Hamiltonian $|\phi_i\rangle$ is expanded in the site basis $\{|n\rangle,\, n = 1, \dots, L\}$, then $|\phi_i\rangle = \sum_n c_{in} |n\rangle$, where $c_{in}$ is the $n$-th component of the $i$-th eigenstate. The box probability for the $k$-th box of the $i$-th eigenstate \[p_k(d) = \sum\limits_{n = (k-1)d + 1}^{kd} |c_{in}|^2, ~{\rm where}~ k = 1, \dots, N_d\] is defined as a suitable measure. Here, $|c_{in}|^2 = |u_{i,n}|^2 + |v_{i,n}|^2$ determines the occupation of the site $n$ for the $i$-th eigenstate and $\{u_{i,n},\, v_{i,n}\}$ are the coefficients in the BdG basis. The generalized fractal dimension $D_q$ of any eigenstate can be calculated from the following relation:
\begin{equation}
D_q = \frac{1}{q-1}\lim_{d \rightarrow 0} \frac{\log\left(\sum\limits_{k=1}^{N_d} [p_k(d)]^q\right)}{\log d}.
\label{D_q_eq}
\end{equation}
For the finite dimensional case, $D_q$ is obtained from the slope of a curve, where the numerator in Eq. \eqref{D_q_eq} is plotted as a function of the denominator. The fractal dimension $D_q \simeq 1.0\, (\simeq 0.0)$ for the delocalized (localized) eigenstates. If the dimension $D_q$ of an eigenstate is independent of $q$, then the state is a mono-fractal or simply a fractal state, as it is described by a single scaling exponent. However, if $D_q$ varies with $q$, then the state is a multifractal. 

Besides the fractal dimension of the individual eigenstates, in this section, we characterize the phase diagram of the system with the help of two global observables: the mean inverse participation ratio (MIPR) and the mean normalized participation ratio (MNPR), where the mean is calculated over all the eigenstates. These two observables are used extensively in literature to investigate the localization transition. Following are the definitions of these observables \cite{PhysRevB.106.024204}:
\begin{equation}
{\rm MIPR} = \frac{1}{2N} \sum_{i=1}^{2N} \sum_{n=1}^N \left(|u_{i,n}|^2 + |v_{i,n}|^2\right)^2
\end{equation}
and
\begin{equation}
{\rm MNPR} = \frac{1}{2N} \sum_{i=1}^{2N} \left[N \sum_{n=1}^N \left(|u_{i,n}|^2 + |v_{i,n}|^2\right)^2\right]^{-1},
\end{equation}
where \[\sum_{n=1}^N \left(|u_{i,n}|^2 + |v_{i,n}|^2\right) = 1\] is the normalization condition for each eigenstate $i$. An important point is to be noted that, in the above definition of the MNPR, we have used a factor $N$ within the square-bracket, which was originally a factor $2N$ in Ref. \cite{PhysRevB.106.024204}. As a consequence, the range of the MNPR is now within $\frac{1}{N}$ and $1$, instead of $\frac{1}{2N}$ and $\frac{1}{2}$ in Ref. \cite{PhysRevB.106.024204}. Both these measures are very useful quantities to probe the localization transition of the eigenstates. Here, we again set the system size $N=1597$ while varying the box-size $d$ between $2$ to $20$ for the calculation of $D_2 = D_{q=2}$ for the individual eigenstates. The same values of $d$ are also considered in the calculation of the generalized fractal dimension $D_q$ as a function of $q$.

In Fig. \ref{localization}, we present the results supporting the localization properties of the eigenstates for both short and long-range pairing. Figure  \ref{localization}(a) presents the eigenspectra as a function of the non-Hermitian parameter $\mu$ for the short-range pairing, where $D_2$ is shown in color. This figure shows that all the eigenstates are delocalized for $\mu < 0.38$, and correspondingly $D_2 \simeq 1$. The eigenstates are critical for $0.38<\mu<0.59$, where $D_2$ shows fractional values. For $\mu>0.59$, $D_2$ approaches zero, and the states are localized. The transition point are shown by the blue-colored dashed lines. Furthermore, we present the MIPR and MNPR in Fig. \ref{localization}(b). We expect a prominent transition in MIPR (MNPR) as the system undergoes a phase change. By observing the behavior of the MIPR and MNPR, we observe that the states are delocalized for $\mu<0.38$. Subsequently, MIPR (MNPR) showed an increase (decrease) for $\mu<0.59$ corresponding to the (multi)fractal region. After $\mu = 0.59$, we observe that the MIPR (MNPR) reaches close to its highest (lowest) value, corresponding to the localized region. The black dashed lines at $\mu = 0.38$ and $\mu  = 0.59$ respectively mark the boundary of the delocalized-(multi)fractal and (multi)fractal-localized region. Here we have used pink, orange, and light-green regions to mark respectively the delocalized, intermediate (multi)fractal, and localized regimes on the basis of the behavior of $D_2$. Here, one should note the following important point: in the delocalized and the localized regimes, the fractal dimension $D_2$ of each eigenstate is almost equal, i.e., $D_2 \simeq 1.0$ in the delocalized and $D_2 \simeq 0.0$ in the localized regime. However, in the intermediate (multi)fractal regime, for each value of $\mu$, all the eigenstates are (multi)fractal with significantly different fractional values of $D_2$. Hence, by the orange color, we are indicating the region where $D_2$ is highly fluctuating over all the eigenstates.

We now select three distinct $\mu$ values within each respective region marked by arrows in Fig. \ref{localization}(b). In Fig. \ref{localization}(c), we observe the scaling of the MIPR (open circles) and MNPR (filled circles) with increase in the system size from $N = 34$ to $4181$. For the delocalized states (represented by the red circles), we observe that the MNPR converges to a non-zero value. In contrast, the MIPR follows a decay law characterized by $\sim N^{-1}$, as indicated by the a black dashed line. Conversely, for the localized states (represented by the green circles), the MIPR converges to a non-zero value close to {\it unity}, and this indicates $D_2 \simeq 0$. Here, according to our expectation, the MIPR is independent of the system size. On the other hand, the MNPR decays as $N^{-1}$, shown by the black dashed line. For the (multi)fractal states (represented by orange circles), neither the MIPR nor the MNPR decays as $N^{-1}$. More precisely, they scale as $N^{-D_2}$, with a specific fractional value of $D_2$. 

In Fig. \ref{localization}(d), for $\mu = 0.2$ [marked by the red arrow in the delocalized region of Fig. \ref{localization}(b)], we observe $D_q \simeq 1$ and is independent of $q$ for both the system sizes $(N=1597,\,4181)$. For $\mu = 0.8$ [marked by the orange arrow in Fig. \ref{localization}(b)], $D_q$ not only varies with $q$ and also shows its strong sensitivity to the system size. This indicate multifractal nature of the state. Finally, for $\mu = 1.5$ [marked by the green arrow in the localized region of Fig. \ref{localization}(b)], following our expectation, the dimension $D_q \simeq 0$ and is independent of $q$. Here again, $D_q$ shows weak dependence on the system size, and thus indicating insulating behavior of the state. 

We further observe in Fig. \ref{localization}(e-h) that the behavior of the eigenvalues and the corresponding eigenstates for the long-range pairing case. Figure  \ref{localization}(e) shows that, for the long-range case, the delocalized regime is in the region $\mu<0.44$, whereas the intermediate (multi)fractal regime is within $0.44<\mu<4.9$, and the localized regime is observed for the larger values of the non-Hermitian parameter $\mu$. Here again, we have used pink, orange, and light-green shaded regions to mark the different regimes based on the fractal dimension $D_2$. As earlier, we are representing a region by orange color, where the fractal dimension $D_2$ is highly fluctuating over all the eigenstates. Again, from each of these regions, we randomly selected a $\mu$ value, marked by arrows, and presented the typical scaling property of the MIPR and MNPR with the dimension $N$. In Fig. \ref{localization}(h), the multifractal behavior is confirmed by presenting the scaling behavior of the generalized fractal dimension $D_q$ with $q$. 

These results show that the delocalization-localization transition via multifractal state occurs at a smaller value of $\mu$ for the case of short-range pairing. This finding is in contrast to the Hermitian case, where only delocalizatin to multifractal transition was observed for the long-range pairing \cite{PhysRevB.106.024204} and also for the long-range hopping \cite{PhysRevLett.123.025301, PhysRevB.103.075124}. Our results also differ from the NHAA model with long-range hopping, where localized states are absent \cite{PhysRevB.107.174205}. Additionally, for the case of short-range pairing, we have observed a confluence of two transitions: the topological transition aligns with the multifractal to localized transition, where both are occurring at $\mu = 0.59$. In contrast, our observations for the long-range $p$-wave pairing reveal a distinct behavior. The topological transition follows the delocalized to multifractal transition, and this occurs for smaller value of $\mu$ than the multifractal to localized transition. This finding contradicts the behavior observed in the NH-AAH model with $p$-wave pairing, where the topological phase transition is consistent with a multifractal to localized transition \cite{PhysRevB.105.024514, xu2020fate}. These contrasting results are attributed to the distinctive effects introduced by the long-range pairing. In addition, it is noteworthy that, for both the pairing ranges, the delocalized region aligns with the real region. This observation is evident when examining the imaginary parts, as illustrated in Fig. \ref{Eigenspectra} (a, b). The exploration and analysis of this alignment will be the focus of our investigation in the next section.  

\section{Unconventional real to complex transition} 
\label{sec6}

\begin{figure}[t]
\centering
\begin{tabular}{c}
\includegraphics[width=1\linewidth]{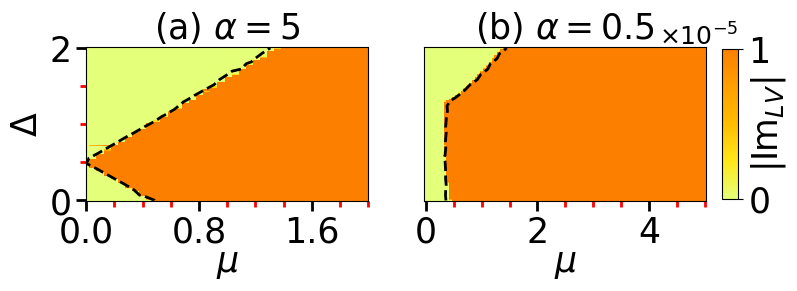} \\
\includegraphics[width=1\linewidth]{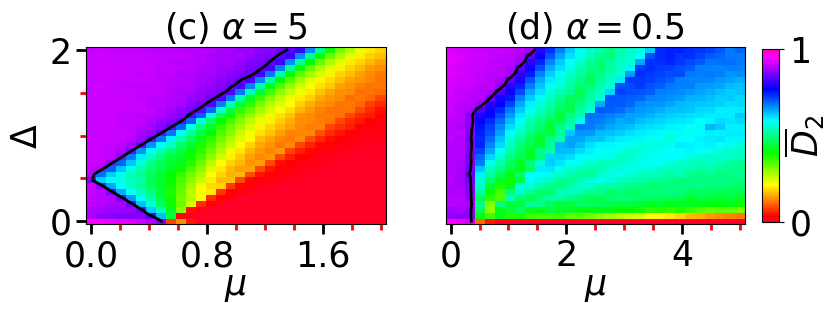} 
\end{tabular}
\caption{The phase diagram illustrates the regions of real eigenvalues, indicated by the yellow color, and imaginary eigenvalues, marked by orange color, as functions of the pairing strength $\Delta$ and the non-Hermitian chemical potential parameter $\mu$ for (a) $\alpha = 5.0$ and (b) $\alpha = 0.5$. The colorbar indicates the maximum value of the imaginary part of the energy eigenvalues. The region of real eigenvalues increases with the increment in the pairing strength $\Delta$. Phase diagrams in subfigures (c) and (d) are respectively showing the the mean fractal dimension $\overline{D}_2$ for different values of $\mu$ and $\Delta$ for the short- and the long-range pairing. The black line marks the boundary between delocalized and multifractal region.}
\label{ImPhase}
\end{figure}

\begin{figure}[b]
\includegraphics[width=1\linewidth]{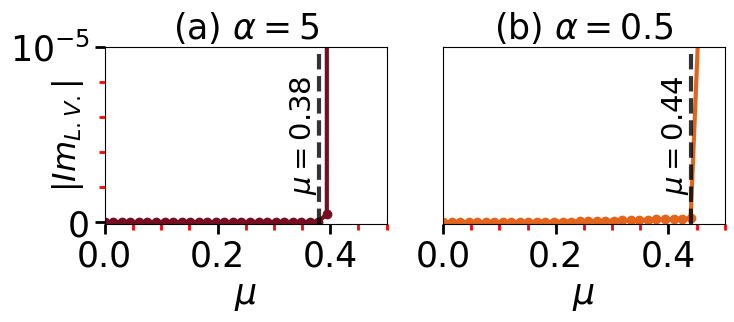}
\caption{The behavior of the largest value of the imaginary part of the energy eigenvalues are presented as a function of the non-Hermitian parameter $\mu$ for the pairing strength $\Delta = 0.1$. Subfigure (a) shows the result for the short-range pairing case with $\alpha = 5.0$, and (b) shows the same for the long-range pairing case with $\alpha = 0.5$. The black dashed lines correspond to the unconventional transition from real to complex eigenvalues.}
\label{Im_mu}
\end{figure}

\begin{figure*}[t]
\centering
\begin{tabular}{cc}

\includegraphics[width=8.3cm,height = 6cm ]{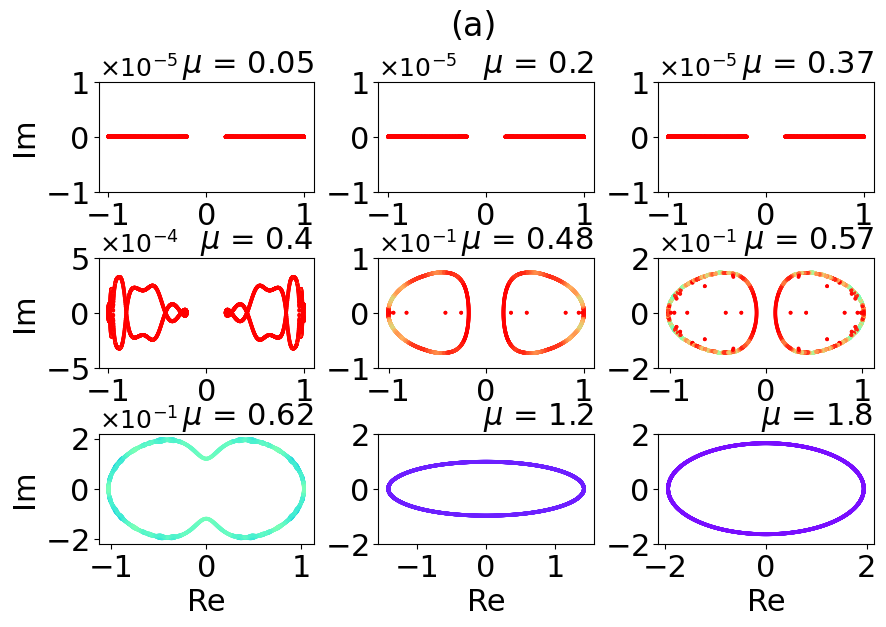} &
\color{red}\vrule
\includegraphics[width=9.5cm, height = 6cm]{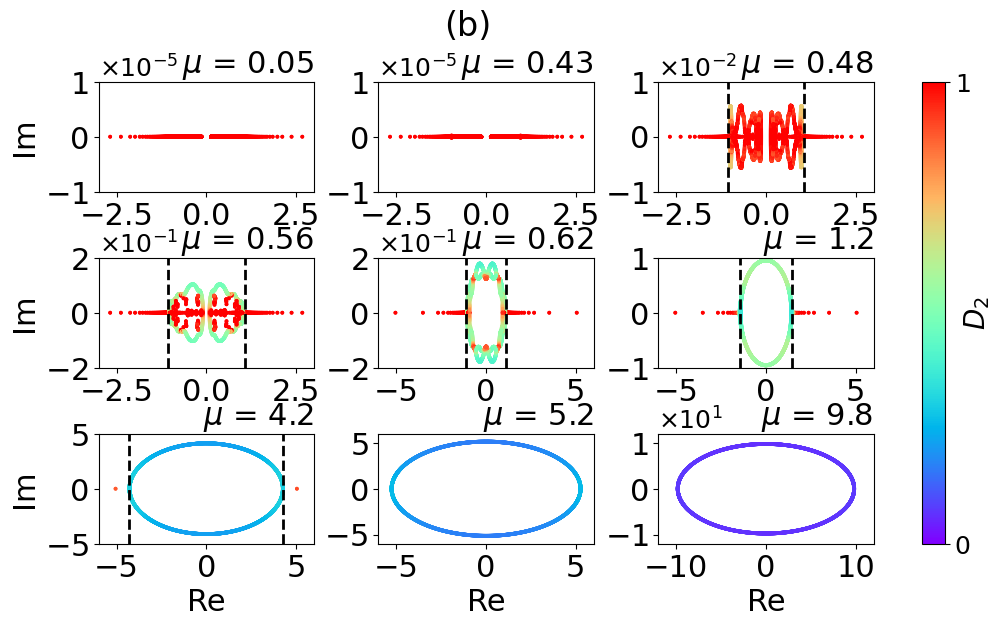} \\
\end{tabular}
\caption{Subfigures (a) and (b) illustrate the real and imaginary parts of the eigenvalues for the short- and long-range pairing for different values of the parameter $\mu$, respectively. The values of $\mu$ are selected from different regimes, as depicted in Fig. \ref{localization}(a) and (e). In subfigure (a), the top row presents the energy eigenvalues at $\mu = 0.05$, $\mu = 0.2$, and $\mu = 0.37$ from the metallic regime. Here, energies are real-valued. The second row includes $\mu = 0.4$, $0.48$, and $0.57$, selected from the multifractal regime. Here, the spectrum exhibits characteristic loops without encircling the origin. The third row shows $\mu = 0.62$, $1.2$, and $1.8$ from the insulating phase, showing a loop encircling the origin in the complex energy plane. The first two boxes of the top row of the subfigure (b) show the energy eigenvalues at $\mu = 0.05$ and $\mu = 0.43$, selected from the metallic regime, where again the energies are real-valued. We choose $\mu = 0.48$, $0.56$, $0.62$, $1.2$, and $4.2$ from the multifractal regime, featuring complex spectra with loops accompanied by points along the real line in the energy plane. The black dashed lines correspond to the mobility edge separating the real energies from the complex ones. We select $\mu = 5.2$ and $9.8$ from the insulating phase, exhibiting loops encircling the origin in the complex energy plane. The color codes of these plots are determined by the corresponding values of the fractal dimension $D_2$, as shown by the color bar.}
\label{Realcomplex}
\end{figure*}

A recent study has reported that a non-PT-symmetric superconducting system (i.e., the Hamiltonian has no PT-symmetry) can also show a real to complex transition of energy \cite{PhysRevB.103.104203}. Upon examining Fig. \ref{Eigenspectra}, we observe a notable transition in the imaginary part of the eigenenergies, signifying an unconventional behavior marked by the shift from real to complex energies. We now study this {\it unconventional} real-complex transition in both short- and long-range pairing scenarios and construct a phase diagram by varying relevant parameters. Additionally, we  the {\it energy spectrum space}, where the real and imaginary parts of the energy eigenvalues are presented to visualize the behavior of the system such as the formation of loops, transition from real to complex energies, and formation of the mobility edges.

\subsection{Phase diagram}

We compute the maximum value of the imaginary part of the energy eigenvalues ($|\mathrm{Im}_{L.V.}|$). These results are presented in the phase diagram shown in Fig. \ref{ImPhase}(a, b). For the numerical simulations, we set the phase $\theta=0$. It is important to note that this transition does not depend on the values of the parameter $\theta$. We observe that the system exhibits a region of real eigenvalues shown by yellow color. In addition, we  provide the mean fractal dimension ($\overline{D}_2$) calculated while tuning $\mu$ and $\Delta$ for both short- and long-range pairing scenarios, as depicted in Fig. \ref{ImPhase} (c, d). The $\overline{D}_2$ is calculated using the formula:
\begin{equation}
\overline{D}_2 = \frac{1}{2N} \sum_{n=1}^{N} D_2(n) .
\end{equation}

We observe that the real region coincides with the parameter regime where delocalized states are observed (pink region in Fig. \ref{ImPhase} (c, d)). This observation suggests a strong correlation between the real and delocalized states in the system for any parameter range, $\mu$ and $\Delta$, for any range of pairing. Further, we observe that with the increase in pairing strength $\Delta$, the region of real eigenvalues increases, which can be attributed to particle-hole symmetry \cite{PhysRevB.103.214202}. Figure \ref{Im_mu} (a, b) shows the behavior of $|\mathrm{Im}_{L.V.}|$ againt $\mu$ for $\Delta = 0.1$. The transition from real to complex eigenenergies is shown by black dashed lines, which coincides with delocalized to multifractal transition as shown in Fig. \ref{localization}(a, e). 

\subsection{Energy spectrum space}

We analyze the energy spectrum space of the Hamiltonian by examining the real and imaginary parts of the eigenvalues and their variation with respect to the system parameters. The result for the short-range pairing case is presented in Fig. \ref{Realcomplex}(a). Here, we have chosen nine different values of the parameter $\mu$ with $\Delta = 0.1$: three values of $\mu$ from a specific region, where two values are chosen near the transition boundaries of that region and one is selected from its interior. We have chosen $\mu = 0.05$, $0.2$ and $0.37$, from the delocalized region, shown by red color in Fig. \ref{localization}(b). Additionally, $\mu = 0.4$, $\mu = 0.48$, and $\mu = 0.57$ are chosen from the orange (multifractal) region, while the remaining three values, $\mu = 0.62$, $\mu = 1.2$, and $\mu = 1.8$, are selected from the green (localized) colored regime. 

The top three subfigures of Fig. \ref{Realcomplex}(a) show that, for all the three values of $\mu$, the eigenvalues form a cantor set on real axis with a gap at origin. However, as we move towards the multifractal region, the eigenvalues start to form loops that do not encircle the origin, once again accompanied by a gap at the origin. This behavior is an indication of the unconventional real to complex transition in energy. As we know from Figs. \ref{Eigenspectra}(a) and \ref{winding}(a), for $\mu< 0.59$, there is a central energy gap and the states exhibit topological non-trivial behavior. Therefore, the presence of a gap at the origin, observed for $\mu < 0.59$, indicates the existence of topological non-trivial states. Furthermore, for the localized regime, we observe a loop structure around the origin formed by the eigenvalues. The color bar shows the values of $D_2$.

For the long-range pairing case, the values of the non-Hermitian parameter $\mu$ are selected based on the metal-insulator behavior described in Fig. \ref{localization}(e). Again, we have chosen nine values of $\mu$. For the delocalized and localized regions, we have selected two values near the boundary of transition, respectively. Whereas for the multifractal region, we have chosen five values, where two values are chosen from $\mu = (0.44, 0.59)$, representing the topological non-trivial region, and the remaining three values are chosen from the topological trivial region $\mu = (0.59, 4.9)$.

In Fig. \ref{Realcomplex}(b), it is evident that, initially, the energy eigenvalues remain real with a central gap for $\mu = 0.05$ and $0.43$. However, for $\mu = 0.48$ and $0.56$, the system has multifractal states. At these two values of $\mu$, the eigenvalues become complex and form multiple loops. These loops do not encircle the origin, but exhibit a gap at the origin. This observation is consistent with the presence of a central gap for $\mu < 0.59$, as depicted in Figs. \ref{Eigenspectra}(b) and \ref{winding}(b), signifying a topologically non-trivial region. These loops are accompanied by the points on the real line outside the respective intervals and these are delineated by the black dashed vertical lines. Specifically, for $\mu = 0.48$, the interval is $[-1.003, \, 1.003]$, and for $\mu = 0.56$, it is $[-1.03,\, 1.03]$. Remarkably, at the parameter values $\mu = 0.62,\, 1.2,$ and $4.2$, the eigenvalues form loops that enclose the origin within the real energy intervals $[-1.07,\, 1.07],\, [-1.46,\, 1.46],$ and $[-4.28,\, 4.28]$, respectively, and these are marked by the black dashed lines. Again, these loops are accompanied by points on the real line outside their respective intervals. Here also the color of the energy eigenvalues indicate the fractal dimension $D_2$ of the corresponding eigenstates. We observe that the complex energies are associated with the localized states, while the real energies are associated with the delocalized states. The black dashed-lines denote the mobility edges, which are effectively separating the real energies from the complex ones. This pattern in the mobility edges is observed for other $\mu$ values within the multifractal regime. 

The following important point also to be noted for the long-range pairing case with multifractal eigenstates. For the nontrivial topological cases, where $\mu = 0.48$ and $0.56$, we do not see any loop encircling the origin. However, for the topologically trivial cases with $\mu=0.62,\,1.2,$ and $4.2$, we observe a loop encircling the origin accompanied by points along the real line. The last two subplots depict eigenvalues forming a loop encircling the origin, which is indicating the presence of only localized states within the system and there is no mobility edge. 

The described loop structure formation in the energy spectrum space due to the unconventional real-complex transition has been observed in various systems. In superconducting systems with nearest-neighbor pairing \cite{PhysRevB.105.024514}, similar loop structures have been reported during the transition from real to complex energies. However, the long-range pairing leads to the formation of mobility edge, which contradicts the behavior observed in previous studies. Furthermore, we infer from the results obtained from the selected values of $\mu$ that, in this model, the unconventional real to complex transition  and the metallic to multifractal transition occur at the same value of the parameter $\mu$.

\section{Summary}
\label{sec7}

Our investigation of the non-Hermitian AA chain with $p$-wave pairing has yielded profound insights into its behavior under short- and long-range pairing cases. Notably, we have observed two distinct phenomena: the emergence of MZMs under short-range pairing and MDMs under long-range pairing. Particularly intriguing is the occurrence of a significant crossing at zero energy in the MDMs under long-range pairing, establishing an adiabatic connection with MZMs, dependent on the phase value (see Fig. \ref{Eigenspectra}). Additionally, we have identified simultaneous topological and superconducting phase transitions across both pairing ranges. The MZMs exhibit a winding number of $\nu = 1$, while for MDMs, $\nu = \frac{1}{2}$.

Furthermore, our investigation into localization properties has unveiled delocalized, multifractal, and localized regions. Remarkably, we have observed that the topological phase transition coincides with the multifractal-to-localized phase transition in the case of short-range pairing. Hence, we do not see any loop in the energy spectrum space encircling the origin. On the other hand, for the long-range pairing case, the topological phase transition precedes the multifractal-to-localized phase transition. As a consequence, we observe loops, which do not encircle the origin for the topologically nontrivial cases. For the trivial cases, we observe mobility edges in the energy spectrum space, which separate the real energy region from the loop encircling the origin. In topological non-trivial case also, we have mobility edge that separates the real energy from the loop not encircling the origin. Intriguingly, due to the particle-hole symmetry in the system, we have discovered an unconventional real-complex transition of eigenenergies. From the phase diagrams, we infer that the observed unconventional real-to-complex transition in this model coincides with the delocalized to multifractal transition for all $\mu$ and $\Delta$ values. In the delocalized phase, energies are entirely real. These findings underscore the rich and diverse phenomena that arise in non-Hermitian systems with different types of pairing, providing valuable insights into the behavior of such systems.

\begin{acknowledgments}
One of the authors (JNB) acknowledges financial support from DST-SERB, India through a Core Research Grant CRG/2020/001701 and also through a MATRICS grant MTR/2022/000691. Authors also thank the anonymous referees for their extensive and valuable comments, which was immensely helpful to improve the quality of this paper. 

\end{acknowledgments}

\appendix

\section{Details of the  Hamiltonian formulation in the BdG basis}
\label{appendixA}

\subsection{$H_{\rm local}$, $H_{\rm hop}$ and $H_{\rm pair}^{(l)}$ in Eq.   \eqref{bdgHam} }
\label{appendixA1}

\begin{figure*}
\includegraphics[width=0.9\textwidth]{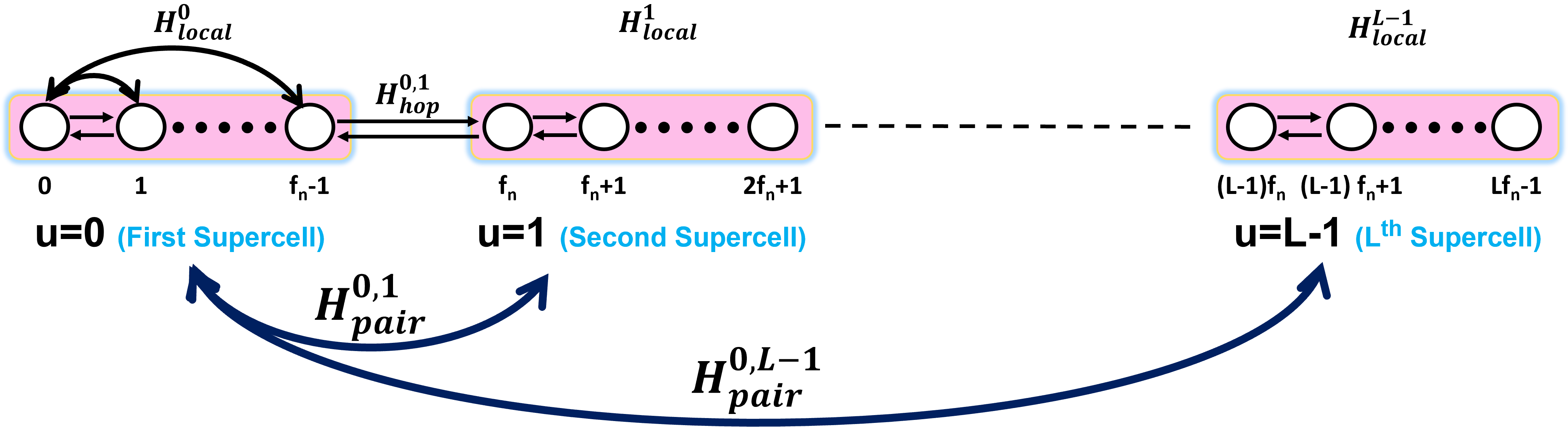}
\caption{We present a schematic diagram of a one-dimensional non-Hermitian Aubry-Andr\'e-Harper (AAH) model. Here, the white circles represent lattice sites, each featuring an onsite complex chemical potential. The system comprises a total of $N = Lf_n$ sites, with $L$ being the total number of supercells and each supercell containing $f_n$ sites. The term $H_{\rm local}$ accounts for interactions within each individual supercell, with the superscript indicating the supercell index. Notably, within the first supercell, curved lines denote pairing for the first site with the remaining sites within the supercell, while double arrows illustrate nearest-neighbor hopping interactions. Additionally, $H_{\rm hop}^{0,1}$ describes hopping between the last lattice site of the first supercell ($u = 0$) and the first lattice site of the second supercell ($u = 1$). $H_{\rm pair}^{0,1}$, marked by a blue curved line, represents pairing between the first and the second supercell, while $H_{\rm pair}^{0,L-1}$ signifies pairing between the first and the last supercell.}
\label{model}
\end{figure*}

We present the explicit expressions for the components of the BdG Hamiltonian, as defined in Eq. \eqref{bdgHam}. These components include the local term ($H_{\rm local}$), which represents interactions within individual supercells; the hopping term ($H_{\rm hop}$), responsible for interactions between all supercells, and the long-range pairing term ($H_{\rm pair}^{(l)}$), which establishes connections across all supercells through long-range superconducting pairing. For a visual representation of these terms, refer to Fig. \ref{model}. 
\begin{equation}
H_{\rm local} = 
\begin{pmatrix}
A_0 & B & C_2 & \cdots & C_{f_n-1} \\
B^{\dagger} & A_1 & B & \cdots & C_{f_n-2} \\
\vdots & \vdots & \vdots & \ddots & \vdots \\
C^{\dagger}_{f_n-2} & C^{\dagger}_{f_n-3} & C^{\dagger}_{f_n-4} & \cdots & B \\
C^{\dagger}_{f_n-1} & C^{\dagger}_{f_n-2} & C^{\dagger}_{f_n-3} & \cdots & A_{f_n-1} \\
\end{pmatrix},
\label{A1}
\end{equation}
\begin{equation}
H_{\rm hop} = 
\begin{pmatrix}
0 & 0 & 0 & \cdots & 0 \\
0 & 0 & 0 & \cdots & 0 \\
\vdots & \vdots & \vdots & \ddots & \vdots \\
B^\prime & 0 & 0 & \cdots & 0 \\
\end{pmatrix},
\label{A2}
\end{equation}
and 
\begin{equation}
H_{\rm pair}^{(l)} = 
\begin{pmatrix}
C_{l,0,0} & C_{l,0,1} & C_{l,0,2} & \cdots & C_{l,0,f_n-1} \\
C_{l,1,0} & C_{l,1,1} & C_{l,1,2} & \cdots & C_{l,1,f_n-1} \\
\vdots & \vdots & \vdots & \ddots & \vdots \\
C_{l,f_n-2,0} & C_{l,f_n-2,1} & C_{l,f_n-2,2} & \cdots & C_{l,f_n-2,f_n-1} \\
C_{l,f_n-1,0} & C_{l,f_n-1,1} & C_{l,f_n-1,2} & \cdots & C_{l,f_n-1,f_n-1} \\
\end{pmatrix}
\label{A3}
\end{equation}
 where $A_j = -\mu f(j) \sigma_{z}$, $B = \frac{t}{2} \sigma_{z} - \Delta i \sigma_{y}$, $C_l = -\frac{\Delta}{d_l^{\alpha} i \sigma_y}$, $B' = \frac{t}{2} \sigma_{z}$ and $C_{l,x,y} =  \frac{\Delta}{2d_{l,x,y}^{\alpha}}i \sigma_{y}$. For the systems with APBC $d_l = \min (l,f_n-l)$ and $d_{l,x,y} = \min (lf_n - x + y, f_n-lf_n + x-y)$.
 
\subsection{$H^{\rm BdG}$ for $L = 1$ supercell}
\label{appendixA2}

Here, we find the expression for the BdG Hamiltonian given in Eq. \eqref{bdgHam} for a single supercell, specifically when $L=1$. For this case, BdG basis can be writen as
$\chi = (c_{0}, c_{0}^{\dagger}, \ldots, c_{N-1}, c_{N-1}^{\dagger})^{T}$. Therefore, the Hamiltonian described by Eq. \eqref{1} can be written as follows:
\begin{equation}
 H = \chi^{\dagger} H^{\rm BdG}\chi
\end{equation}  
where
\begin{equation}
H^{\rm BdG} = 
\begin{pmatrix}
A_0 & B & C_2 & \cdots & -B^{\dagger} \\
B^{\dagger} & A_1 & B & \cdots & C_{f_n-2} \\
\vdots & \vdots & \vdots & \ddots & \vdots \\
C^{\dagger}_{f_n-2} & C^{\dagger}_{f_n-3} & C^{\dagger}_{f_n-4} & \cdots & B \\
-B & C^{\dagger}_{f_n-2} & C^{\dagger}_{f_n-3} & \cdots & A_{f_n-1} \\
\end{pmatrix}
\label{2}
\end{equation}
This Hamiltonian $H^{\rm BdG}$ is a $2f_n \times 2f_n$ matrix with $A_i = -Vf(i)\sigma_{z}$, $B = \frac{t}{2}\sigma_z - \Delta i \sigma_{y}$, and $C_l = -\frac{\Delta}{{d_{l}}^{\alpha}}i\sigma_{y}$. Here, we replace $l$ in Eq. \eqref{1} with $d_l = \min(l, f_n - l)$, because we impose anti-periodic boundary conditions (APBCs) \cite{PhysRevResearch.3.013148}. For the systems with OBC $d_l = l$ and we drop the terms containing $c_{j}>N-1$.  

\section{$Eigenspectra for \Delta = 0.5$}
\label{appendixB}

\begin{figure}[h]

    \includegraphics[width=0.48\textwidth]{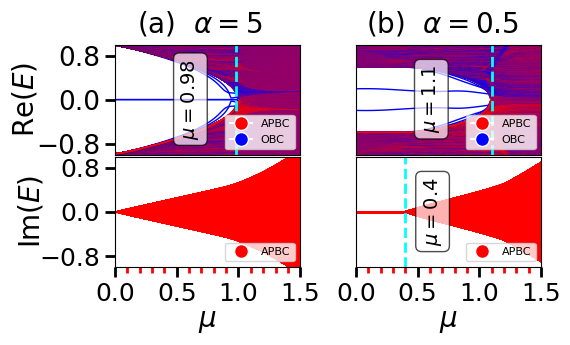} 

\caption{The real and imaginary parts of eigenvalues for the system with OBC (blue) and APBC (red) for the short-range ($\alpha = 5.0$) and the long-range pairing ($\alpha = 0.5$) are shown as a function of the parameter $\mu$. The other system are set as $t=1$, $\Delta = 0.5t$, and $\theta = 0$. (a) The observation of real eigenvalues indicates the emergence of MZMs in the central region $0 \leq \mu \lesssim 0.98$. However, after inspecting the imaginary part of the eigenspectra, we find that no eigenvalues possess real values at these particular parameter values. (b) We observe appearance of MDMs in the central region $0 \leq \mu \lesssim 1.1$ as evidenced by the observation of real eigenvalues. However, the imaginary part of the eigenspectra reveals that the eigenvalues are real for $0 \leq \mu \lesssim 0.4$. }

  \label{B}
\end{figure}

As discussed in the main text, we observe the coexistence of MZMs and MDMs within the same parameter region, for the chosen value of $\Delta = 0.1$. Here, we further investigate this phenomenon and examine that it is not true for other values of $\Delta$. Specifically, we consider the case where $\Delta = 0.5$ and demonstrate that the occurrence of MZMs and MDMs does not overlap within the same parameter regime. From Fig. \ref{B}, we observe that MZMs appear for $\mu<0.98$ while MDMs appear for $\mu<1.1$.

\section{Momentum Representation and Thermodynamic Limit of $H^{\rm BdG}$}\label{appendixC}

In this appendix, we delve into the momentum representation of $H^{\rm BdG}$, and explore its behavior in the thermodynamic limit. 

\subsection{Fourier transform of $H^{\rm BdG}$}
\label{appendixC1}

We explicitly write the BdG Hamiltonian given in Eq. \eqref{bdgHam} as
\begin{equation}
\begin{aligned}
H^{\rm BdG} = \sum_{u=0}^{L-1} \big[ \chi_u^{\dagger} H_{\rm local} \chi_u+ \chi_u^{\dagger} H_{\rm hop} \chi_{u+1} \\
+ \sum_{l=1}^{L-1} ( \chi_u^{\dagger} H_{\rm pair}^{(l)} \chi_{u+l} + {\rm h.c.}) \big]
\end{aligned}
\label{C1}
\end{equation}
where $u \in {0, 1, \ldots, L-1}$ is a supercell index.
Using Fourier transform
\begin{equation}
\chi_{u} = \frac{1}{\sqrt{L}} \sum_{k} e^{i k u} \chi_{k}
\label{C2}
\end{equation}
where $\chi_k = (c_{k,0}, c_{-k,0}^{\dagger}, \ldots, c_{k ,f_n -1}, c_{-k + f_n-1}^{\dagger})^{T}$. Further, we impose APBC by assuming 
\begin{equation}
\chi_{u+L} = -\chi_{u}
\label{C3}
\end{equation}
Using Eq.  \eqref{C2} and Eq.  \eqref{C3}, we get
\begin{equation}
\frac{1}{\sqrt{L}} \sum_{k} e^{i k (u+L)} \chi_{k} = -\frac{1}{\sqrt{L}} \sum_{k} e^{i k u} \chi_{k}
\end{equation}
$k= \frac{(2m+1)\pi}{L}$, $m=0,1, \ldots,L-1$
Thus, the Hamiltonian in momentum representation can be written as
\begin{equation}
\begin{aligned}
H^{\rm BdG} &= \sum_{k} \big[ \chi_k^{\dagger} H_{\rm local} \chi_k+ (e^{ik}\chi_k^{\dagger} H_{\rm hop} \chi_{k}+ {\rm h.c}) \\
&+ \sum_{l=1}^{L-1} ( e^{ikl}\chi_k^{\dagger} H_{\rm pair}^{(l)} \chi_{k} + {h.c.}) \big]
\end{aligned}
\label{C4}
\end{equation}

\subsection{$H^{\rm BdG}$ for the thermodynamic limit $L\rightarrow\infty$}
\label{appendixC2}

For $L\rightarrow\infty$, the Hamiltonian  in Eq. \eqref{C4} reads as 
\begin{equation}
\begin{aligned}
H_{\rm inf} = \sum_{k} \big[ \chi_k^{\dagger} H_{\rm local} \chi_k+ (e^{ik}\chi_k^{\dagger} H_{\rm hop} \chi_{k}+ {\rm h.c}) \\
+ \chi_k^{\dagger} H_{\rm pair}^{(l^\prime)} \chi_{k} + {\rm h.c.}) \big]
\end{aligned}
\label{C5}
\end{equation}
where 
\begin{equation}
H_{\rm pair}^{(l^\prime)} = \sum_{l=1}^{\infty} e^{ikl} H_{\rm pair}^{(l)} = \sum_{l=1}^{\infty} e^{ikl} \left(\frac{-\Delta}{d_{l,x,y}^\alpha} i \sigma_y\right)
\label{C6}
\end{equation}
Now, using the definition of $d_{l,x,y}$ from Appendix A.1 and putting the limit $L\rightarrow\infty$, we get $d_{l,x,y}=lf_n - (x-y)$. Therefore,
\begin{equation}
H_{\rm pair}^{(l^\prime)} = \sum_{l=1}^{\infty} e^{ikl}  \left(\frac{-\Delta}{[lf_n - (x-y)]^\alpha} i \sigma_y\right) = ig_{xy}\sigma_{y} = C_{xy}
\label{C7}
\end{equation}
where 
\begin{equation}
\begin{aligned}
g_{xy} &= -\sum_{l=1}^{\infty} e^{ikl} \left(\frac{\Delta}{[lf_n - (x-y)]^\alpha}\right) \\
&= \frac{-\Delta e^{ik}}{f_n^\alpha} \sum_{l=0}^{\infty} \frac{e^{ikl}}{\left[l+ \frac{f_n-(x-y)}{f_n}\right]^\alpha}
\end{aligned}
\label{C8}
\end{equation}
The form of $g_{xy}$ can be easily obtained by substituting $l\rightarrow{l+1}$, so that the sum starts from $l=0$. Also, we can write $g_{xy}$ in the form of 
\begin{equation}
\begin{aligned}
g_{xy} 
&= -\frac{\Delta e^{ik}}{f_n^\alpha} HLP_{\alpha}(k,f_n,x-y)
\end{aligned}
\label{C9}
\end{equation}
where  $HLP_{\alpha}(k,f_n,x-y) =
\sum_{l=0}^{\infty} \frac{e^{ikl}}{\left[l+ \frac{f_n-(x-y)}{f_n}\right]^\alpha}$ is known as Hurwitz-Lerch-Phi function. The HLP function has a singularity at $k = 0$ for $\alpha<1$. Thus, the form of the matrix $H_{\rm pair}^{(l^\prime)}$ is
\begin{equation}
H_{\rm pair}^{(l^\prime)} = 
\begin{pmatrix}
C_{0,0} & C_{0,1} & C_{0,2} & \cdots & C_{0,f_n-1} \\
C_{1,0} & C_{1,1} & C_{1,2} & \cdots & C_{1,f_n-1} \\
\vdots & \vdots & \vdots & \ddots & \vdots \\
C_{f_n-2,0} & C_{f_n-2,1} & C_{f_n-2,2} & \cdots & C_{f_n-2,f_n-1} \\
C_{f_n-1,0} & C_{f_n-1,1} & C_{f_n-1,2} & \cdots & C_{f_n-1,f_n-1} \\
\end{pmatrix}
\label{C6}
\end{equation}.


\begin{thebibliography}{59}%
\makeatletter
\providecommand \@ifxundefined [1]{%
 \@ifx{#1\undefined}
}%
\providecommand \@ifnum [1]{%
 \ifnum #1\expandafter \@firstoftwo
 \else \expandafter \@secondoftwo
 \fi
}%
\providecommand \@ifx [1]{%
 \ifx #1\expandafter \@firstoftwo
 \else \expandafter \@secondoftwo
 \fi
}%
\providecommand \natexlab [1]{#1}%
\providecommand \enquote  [1]{``#1''}%
\providecommand \bibnamefont  [1]{#1}%
\providecommand \bibfnamefont [1]{#1}%
\providecommand \citenamefont [1]{#1}%
\providecommand \href@noop [0]{\@secondoftwo}%
\providecommand \href [0]{\begingroup \@sanitize@url \@href}%
\providecommand \@href[1]{\@@startlink{#1}\@@href}%
\providecommand \@@href[1]{\endgroup#1\@@endlink}%
\providecommand \@sanitize@url [0]{\catcode `\\12\catcode `\$12\catcode
  `\&12\catcode `\#12\catcode `\^12\catcode `\_12\catcode `\%12\relax}%
\providecommand \@@startlink[1]{}%
\providecommand \@@endlink[0]{}%
\providecommand \url  [0]{\begingroup\@sanitize@url \@url }%
\providecommand \@url [1]{\endgroup\@href {#1}{\urlprefix }}%
\providecommand \urlprefix  [0]{URL }%
\providecommand \Eprint [0]{\href }%
\providecommand \doibase [0]{https://doi.org/}%
\providecommand \selectlanguage [0]{\@gobble}%
\providecommand \bibinfo  [0]{\@secondoftwo}%
\providecommand \bibfield  [0]{\@secondoftwo}%
\providecommand \translation [1]{[#1]}%
\providecommand \BibitemOpen [0]{}%
\providecommand \bibitemStop [0]{}%
\providecommand \bibitemNoStop [0]{.\EOS\space}%
\providecommand \EOS [0]{\spacefactor3000\relax}%
\providecommand \BibitemShut  [1]{\csname bibitem#1\endcsname}%
\let\auto@bib@innerbib\@empty

\bibitem [{\citenamefont {Serge~Aubry}(1980)}]{AubryAndre}%
  \BibitemOpen
  \bibfield  {author} {\bibinfo {author} {\bibfnamefont {G.~A.}\ \bibnamefont
  {Serge~Aubry}},\ }\bibfield  {title} {\bibinfo {title} {Analyticity breaking
  and anderson localization in incommensurate lattices},\ }\bibfield  {journal}
  {\bibinfo  {journal} {Annals of the Israel Physical society}\ }\href
  {https://doi.org///chaos.if.uj.edu.pl/~delande/Lectures/files/An.Is.Phys.Soc.pdf}
  {//chaos.if.uj.edu.pl/~delande/Lectures/files/An.Is.Phys.Soc.pdf} (\bibinfo
  {year} {1980})\BibitemShut {NoStop}%
\bibitem [{\citenamefont {Harper}(1955)}]{Harper}%
  \BibitemOpen
  \bibfield  {author} {\bibinfo {author} {\bibfnamefont {P.~G.}\ \bibnamefont
  {Harper}},\ }\bibfield  {title} {\bibinfo {title} {Single band motion of
  conduction electrons in a uniform magnetic field},\ }\href
  {https://doi.org/10.1088/0370-1298/68/10/304} {\bibfield  {journal} {\bibinfo
   {journal} {Proceedings of the Physical Society A}\ }\textbf {\bibinfo
  {volume} {68}},\ \bibinfo {pages} {874} (\bibinfo {year} {1955})}\BibitemShut
  {NoStop}%
\bibitem [{\citenamefont {Kitaev}(2001)}]{A_Yu_Kitaev_2001}%
  \BibitemOpen
  \bibfield  {author} {\bibinfo {author} {\bibfnamefont {A.~Y.}\ \bibnamefont
  {Kitaev}},\ }\bibfield  {title} {\bibinfo {title} {Unpaired majorana fermions
  in quantum wires},\ }\href {https://doi.org/10.1070/1063-7869/44/10S/S29}
  {\bibfield  {journal} {\bibinfo  {journal} {Physics-Uspekhi}\ }\textbf
  {\bibinfo {volume} {44}},\ \bibinfo {pages} {131} (\bibinfo {year}
  {2001})}\BibitemShut {NoStop}%
\bibitem [{\citenamefont {Chang}\ \emph {et~al.}(1997)\citenamefont {Chang},
  \citenamefont {Ikezawa},\ and\ \citenamefont {Kohmoto}}]{PhysRevB.55.12971}%
  \BibitemOpen
  \bibfield  {author} {\bibinfo {author} {\bibfnamefont {I.}~\bibnamefont
  {Chang}}, \bibinfo {author} {\bibfnamefont {K.}~\bibnamefont {Ikezawa}},\
  and\ \bibinfo {author} {\bibfnamefont {M.}~\bibnamefont {Kohmoto}},\
  }\bibfield  {title} {\bibinfo {title} {Multifractal properties of the wave
  functions of the square-lattice tight-binding model with
  next-nearest-neighbor hopping in a magnetic field},\ }\href
  {https://doi.org/10.1103/PhysRevB.55.12971} {\bibfield  {journal} {\bibinfo
  {journal} {Phys. Rev. B}\ }\textbf {\bibinfo {volume} {55}},\ \bibinfo
  {pages} {12971} (\bibinfo {year} {1997})}\BibitemShut {NoStop}%
\bibitem [{\citenamefont {Liu}\ \emph {et~al.}(2015)\citenamefont {Liu},
  \citenamefont {Ghosh},\ and\ \citenamefont {Chong}}]{PhysRevB.91.014108}%
  \BibitemOpen
  \bibfield  {author} {\bibinfo {author} {\bibfnamefont {F.}~\bibnamefont
  {Liu}}, \bibinfo {author} {\bibfnamefont {S.}~\bibnamefont {Ghosh}},\ and\
  \bibinfo {author} {\bibfnamefont {Y.~D.}\ \bibnamefont {Chong}},\ }\bibfield
  {title} {\bibinfo {title} {Localization and adiabatic pumping in a
  generalized aubry-andr\'e-harper model},\ }\href
  {https://doi.org/10.1103/PhysRevB.91.014108} {\bibfield  {journal} {\bibinfo
  {journal} {Phys. Rev. B}\ }\textbf {\bibinfo {volume} {91}},\ \bibinfo
  {pages} {014108} (\bibinfo {year} {2015})}\BibitemShut {NoStop}%
\bibitem [{\citenamefont {Nayak}\ \emph {et~al.}(2008)\citenamefont {Nayak},
  \citenamefont {Simon}, \citenamefont {Stern}, \citenamefont {Freedman},\ and\
  \citenamefont {Das~Sarma}}]{RevModPhys.80.1083}%
  \BibitemOpen
  \bibfield  {author} {\bibinfo {author} {\bibfnamefont {C.}~\bibnamefont
  {Nayak}}, \bibinfo {author} {\bibfnamefont {S.~H.}\ \bibnamefont {Simon}},
  \bibinfo {author} {\bibfnamefont {A.}~\bibnamefont {Stern}}, \bibinfo
  {author} {\bibfnamefont {M.}~\bibnamefont {Freedman}},\ and\ \bibinfo
  {author} {\bibfnamefont {S.}~\bibnamefont {Das~Sarma}},\ }\bibfield  {title}
  {\bibinfo {title} {Non-abelian anyons and topological quantum computation},\
  }\href {https://doi.org/10.1103/RevModPhys.80.1083} {\bibfield  {journal}
  {\bibinfo  {journal} {Rev. Mod. Phys.}\ }\textbf {\bibinfo {volume} {80}},\
  \bibinfo {pages} {1083} (\bibinfo {year} {2008})}\BibitemShut {NoStop}%
\bibitem [{\citenamefont {Vodola}\ \emph {et~al.}(2014)\citenamefont {Vodola},
  \citenamefont {Lepori}, \citenamefont {Ercolessi}, \citenamefont {Gorshkov},\
  and\ \citenamefont {Pupillo}}]{PhysRevLett.113.156402}%
  \BibitemOpen
  \bibfield  {author} {\bibinfo {author} {\bibfnamefont {D.}~\bibnamefont
  {Vodola}}, \bibinfo {author} {\bibfnamefont {L.}~\bibnamefont {Lepori}},
  \bibinfo {author} {\bibfnamefont {E.}~\bibnamefont {Ercolessi}}, \bibinfo
  {author} {\bibfnamefont {A.~V.}\ \bibnamefont {Gorshkov}},\ and\ \bibinfo
  {author} {\bibfnamefont {G.}~\bibnamefont {Pupillo}},\ }\bibfield  {title}
  {\bibinfo {title} {Kitaev chains with long-range pairing},\ }\href
  {https://doi.org/10.1103/PhysRevLett.113.156402} {\bibfield  {journal}
  {\bibinfo  {journal} {Phys. Rev. Lett.}\ }\textbf {\bibinfo {volume} {113}},\
  \bibinfo {pages} {156402} (\bibinfo {year} {2014})}\BibitemShut {NoStop}%
\bibitem [{\citenamefont {Vodola}\ \emph {et~al.}(2015)\citenamefont {Vodola},
  \citenamefont {Lepori}, \citenamefont {Ercolessi},\ and\ \citenamefont
  {Pupillo}}]{Vodola_2016}%
  \BibitemOpen
  \bibfield  {author} {\bibinfo {author} {\bibfnamefont {D.}~\bibnamefont
  {Vodola}}, \bibinfo {author} {\bibfnamefont {L.}~\bibnamefont {Lepori}},
  \bibinfo {author} {\bibfnamefont {E.}~\bibnamefont {Ercolessi}},\ and\
  \bibinfo {author} {\bibfnamefont {G.}~\bibnamefont {Pupillo}},\ }\bibfield
  {title} {\bibinfo {title} {Long-range ising and kitaev models: phases,
  correlations and edge modes},\ }\href
  {https://doi.org/10.1088/1367-2630/18/1/015001} {\bibfield  {journal}
  {\bibinfo  {journal} {New Journal of Physics}\ }\textbf {\bibinfo {volume}
  {18}},\ \bibinfo {pages} {015001} (\bibinfo {year} {2015})}\BibitemShut
  {NoStop}%
\bibitem [{\citenamefont {Viyuela}\ \emph {et~al.}(2016)\citenamefont
  {Viyuela}, \citenamefont {Vodola}, \citenamefont {Pupillo},\ and\
  \citenamefont {Martin-Delgado}}]{PhysRevB.94.125121}%
  \BibitemOpen
  \bibfield  {author} {\bibinfo {author} {\bibfnamefont {O.}~\bibnamefont
  {Viyuela}}, \bibinfo {author} {\bibfnamefont {D.}~\bibnamefont {Vodola}},
  \bibinfo {author} {\bibfnamefont {G.}~\bibnamefont {Pupillo}},\ and\ \bibinfo
  {author} {\bibfnamefont {M.~A.}\ \bibnamefont {Martin-Delgado}},\ }\bibfield
  {title} {\bibinfo {title} {Topological massive dirac edge modes and
  long-range superconducting hamiltonians},\ }\href
  {https://doi.org/10.1103/PhysRevB.94.125121} {\bibfield  {journal} {\bibinfo
  {journal} {Phys. Rev. B}\ }\textbf {\bibinfo {volume} {94}},\ \bibinfo
  {pages} {125121} (\bibinfo {year} {2016})}\BibitemShut {NoStop}%
\bibitem [{\citenamefont {Fraxanet}\ \emph {et~al.}(2021)\citenamefont
  {Fraxanet}, \citenamefont {Bhattacharya}, \citenamefont {Grass},
  \citenamefont {Rakshit}, \citenamefont {Lewenstein},\ and\ \citenamefont
  {Dauphin}}]{PhysRevResearch.3.013148}%
  \BibitemOpen
  \bibfield  {author} {\bibinfo {author} {\bibfnamefont {J.}~\bibnamefont
  {Fraxanet}}, \bibinfo {author} {\bibfnamefont {U.}~\bibnamefont
  {Bhattacharya}}, \bibinfo {author} {\bibfnamefont {T.}~\bibnamefont {Grass}},
  \bibinfo {author} {\bibfnamefont {D.}~\bibnamefont {Rakshit}}, \bibinfo
  {author} {\bibfnamefont {M.}~\bibnamefont {Lewenstein}},\ and\ \bibinfo
  {author} {\bibfnamefont {A.}~\bibnamefont {Dauphin}},\ }\bibfield  {title}
  {\bibinfo {title} {Topological properties of the long-range kitaev chain with
  aubry-andr\'e-harper modulation},\ }\href
  {https://doi.org/10.1103/PhysRevResearch.3.013148} {\bibfield  {journal}
  {\bibinfo  {journal} {Phys. Rev. Res.}\ }\textbf {\bibinfo {volume} {3}},\
  \bibinfo {pages} {013148} (\bibinfo {year} {2021})}\BibitemShut {NoStop}%
\bibitem [{\citenamefont {DeGottardi}\ \emph {et~al.}(2013)\citenamefont
  {DeGottardi}, \citenamefont {Sen},\ and\ \citenamefont
  {Vishveshwara}}]{PhysRevLett.110.146404}%
  \BibitemOpen
  \bibfield  {author} {\bibinfo {author} {\bibfnamefont {W.}~\bibnamefont
  {DeGottardi}}, \bibinfo {author} {\bibfnamefont {D.}~\bibnamefont {Sen}},\
  and\ \bibinfo {author} {\bibfnamefont {S.}~\bibnamefont {Vishveshwara}},\
  }\bibfield  {title} {\bibinfo {title} {Majorana fermions in superconducting
  1d systems having periodic, quasiperiodic, and disordered potentials},\
  }\href {https://doi.org/10.1103/PhysRevLett.110.146404} {\bibfield  {journal}
  {\bibinfo  {journal} {Phys. Rev. Lett.}\ }\textbf {\bibinfo {volume} {110}},\
  \bibinfo {pages} {146404} (\bibinfo {year} {2013})}\BibitemShut {NoStop}%
\bibitem [{\citenamefont {Cai}\ \emph {et~al.}(2013)\citenamefont {Cai},
  \citenamefont {Lang}, \citenamefont {Chen},\ and\ \citenamefont
  {Wang}}]{PhysRevLett.110.176403}%
  \BibitemOpen
  \bibfield  {author} {\bibinfo {author} {\bibfnamefont {X.}~\bibnamefont
  {Cai}}, \bibinfo {author} {\bibfnamefont {L.-J.}\ \bibnamefont {Lang}},
  \bibinfo {author} {\bibfnamefont {S.}~\bibnamefont {Chen}},\ and\ \bibinfo
  {author} {\bibfnamefont {Y.}~\bibnamefont {Wang}},\ }\bibfield  {title}
  {\bibinfo {title} {Topological superconductor to anderson localization
  transition in one-dimensional incommensurate lattices},\ }\href
  {https://doi.org/10.1103/PhysRevLett.110.176403} {\bibfield  {journal}
  {\bibinfo  {journal} {Phys. Rev. Lett.}\ }\textbf {\bibinfo {volume} {110}},\
  \bibinfo {pages} {176403} (\bibinfo {year} {2013})}\BibitemShut {NoStop}%
\bibitem [{\citenamefont {Wang}\ \emph {et~al.}(2016)\citenamefont {Wang},
  \citenamefont {Liu}, \citenamefont {Xianlong},\ and\ \citenamefont
  {Hu}}]{PhysRevB.93.104504}%
  \BibitemOpen
  \bibfield  {author} {\bibinfo {author} {\bibfnamefont {J.}~\bibnamefont
  {Wang}}, \bibinfo {author} {\bibfnamefont {X.-J.}\ \bibnamefont {Liu}},
  \bibinfo {author} {\bibfnamefont {G.}~\bibnamefont {Xianlong}},\ and\
  \bibinfo {author} {\bibfnamefont {H.}~\bibnamefont {Hu}},\ }\bibfield
  {title} {\bibinfo {title} {Phase diagram of a non-abelian
  aubry-andr\'e-harper model with $p$-wave superfluidity},\ }\href
  {https://doi.org/10.1103/PhysRevB.93.104504} {\bibfield  {journal} {\bibinfo
  {journal} {Phys. Rev. B}\ }\textbf {\bibinfo {volume} {93}},\ \bibinfo
  {pages} {104504} (\bibinfo {year} {2016})}\BibitemShut {NoStop}%
\bibitem [{\citenamefont {Zeng}\ \emph {et~al.}(2016)\citenamefont {Zeng},
  \citenamefont {Chen},\ and\ \citenamefont {L\"u}}]{PhysRevB.94.125408}%
  \BibitemOpen
  \bibfield  {author} {\bibinfo {author} {\bibfnamefont {Q.-B.}\ \bibnamefont
  {Zeng}}, \bibinfo {author} {\bibfnamefont {S.}~\bibnamefont {Chen}},\ and\
  \bibinfo {author} {\bibfnamefont {R.}~\bibnamefont {L\"u}},\ }\bibfield
  {title} {\bibinfo {title} {Generalized aubry-andr\'e-harper model with
  $p$-wave superconducting pairing},\ }\href
  {https://doi.org/10.1103/PhysRevB.94.125408} {\bibfield  {journal} {\bibinfo
  {journal} {Phys. Rev. B}\ }\textbf {\bibinfo {volume} {94}},\ \bibinfo
  {pages} {125408} (\bibinfo {year} {2016})}\BibitemShut {NoStop}%
\bibitem [{\citenamefont {Xu}\ and\ \citenamefont
  {Li}(2023)}]{10.1093/ptep/ptad043}%
  \BibitemOpen
  \bibfield  {author} {\bibinfo {author} {\bibfnamefont {T.-T.}\ \bibnamefont
  {Xu}}\ and\ \bibinfo {author} {\bibfnamefont {J.-R.}\ \bibnamefont {Li}},\
  }\bibfield  {title} {\bibinfo {title} {{Topological properties in an
  Aubry–André–Harper model with p-wave superconducting pairing}},\
  }\bibfield  {journal} {\bibinfo  {journal} {Progress of Theoretical and
  Experimental Physics}\ }\textbf {\bibinfo {volume} {2023}},\ \href
  {https://doi.org/10.1093/ptep/ptad043} {10.1093/ptep/ptad043} (\bibinfo
  {year} {2023}),\ \bibinfo {note} {043I01},\ \Eprint
  {https://arxiv.org/abs/https://academic.oup.com/ptep/article-pdf/2023/4/043I01/50093911/ptad043.pdf}
  {https://academic.oup.com/ptep/article-pdf/2023/4/043I01/50093911/ptad043.pdf}
  \BibitemShut {NoStop}%
\bibitem [{\citenamefont {Zeng}\ \emph {et~al.}(2021)\citenamefont {Zeng},
  \citenamefont {Lü},\ and\ \citenamefont {You}}]{Zeng_2021}%
  \BibitemOpen
  \bibfield  {author} {\bibinfo {author} {\bibfnamefont {Q.-B.}\ \bibnamefont
  {Zeng}}, \bibinfo {author} {\bibfnamefont {R.}~\bibnamefont {Lü}},\ and\
  \bibinfo {author} {\bibfnamefont {L.}~\bibnamefont {You}},\ }\bibfield
  {title} {\bibinfo {title} {Topological superconductors in one-dimensional
  mosaic lattices},\ }\href {https://doi.org/10.1209/0295-5075/ac1879}
  {\bibfield  {journal} {\bibinfo  {journal} {Europhysics Letters}\ }\textbf
  {\bibinfo {volume} {135}},\ \bibinfo {pages} {17003} (\bibinfo {year}
  {2021})}\BibitemShut {NoStop}%
\bibitem [{\citenamefont {Ashida}\ \emph {et~al.}(2020)\citenamefont {Ashida},
  \citenamefont {Gong},\ and\ \citenamefont {Ueda}}]{ashida2020non}%
  \BibitemOpen
  \bibfield  {author} {\bibinfo {author} {\bibfnamefont {Y.}~\bibnamefont
  {Ashida}}, \bibinfo {author} {\bibfnamefont {Z.}~\bibnamefont {Gong}},\ and\
  \bibinfo {author} {\bibfnamefont {M.}~\bibnamefont {Ueda}},\ }\bibfield
  {title} {\bibinfo {title} {Non-{H}ermitian physics},\ }\href
  {https://doi.org/10.1080/00018732.2021.1876991} {\bibfield  {journal}
  {\bibinfo  {journal} {Advances in Physics}\ }\textbf {\bibinfo {volume}
  {69}},\ \bibinfo {pages} {249} (\bibinfo {year} {2020})},\ \Eprint
  {https://arxiv.org/abs/https://doi.org/10.1080/00018732.2021.1876991}
  {https://doi.org/10.1080/00018732.2021.1876991} \BibitemShut {NoStop}%
\bibitem [{\citenamefont {Bergholtz}\ \emph {et~al.}(2021)\citenamefont
  {Bergholtz}, \citenamefont {Budich},\ and\ \citenamefont
  {Kunst}}]{bergholtz2021exceptional}%
  \BibitemOpen
  \bibfield  {author} {\bibinfo {author} {\bibfnamefont {E.~J.}\ \bibnamefont
  {Bergholtz}}, \bibinfo {author} {\bibfnamefont {J.~C.}\ \bibnamefont
  {Budich}},\ and\ \bibinfo {author} {\bibfnamefont {F.~K.}\ \bibnamefont
  {Kunst}},\ }\bibfield  {title} {\bibinfo {title} {Exceptional topology of
  non-{H}ermitian systems},\ }\href
  {https://doi.org/10.1103/RevModPhys.93.015005} {\bibfield  {journal}
  {\bibinfo  {journal} {Rev. Mod. Phys.}\ }\textbf {\bibinfo {volume} {93}},\
  \bibinfo {pages} {015005} (\bibinfo {year} {2021})}\BibitemShut {NoStop}%
\bibitem [{\citenamefont {Wang}\ \emph {et~al.}(2021)\citenamefont {Wang},
  \citenamefont {Zhang}, \citenamefont {Hua}, \citenamefont {Lei},
  \citenamefont {Lu},\ and\ \citenamefont {Chen}}]{wang2021topological}%
  \BibitemOpen
  \bibfield  {author} {\bibinfo {author} {\bibfnamefont {H.}~\bibnamefont
  {Wang}}, \bibinfo {author} {\bibfnamefont {X.}~\bibnamefont {Zhang}},
  \bibinfo {author} {\bibfnamefont {J.}~\bibnamefont {Hua}}, \bibinfo {author}
  {\bibfnamefont {D.}~\bibnamefont {Lei}}, \bibinfo {author} {\bibfnamefont
  {M.}~\bibnamefont {Lu}},\ and\ \bibinfo {author} {\bibfnamefont
  {Y.}~\bibnamefont {Chen}},\ }\bibfield  {title} {\bibinfo {title}
  {Topological physics of non-{H}ermitian optics and photonics: a review},\
  }\href {https://doi.org/10.1088/2040-8986/ac2e15} {\bibfield  {journal}
  {\bibinfo  {journal} {Journal of Optics}\ }\textbf {\bibinfo {volume} {23}},\
  \bibinfo {pages} {123001} (\bibinfo {year} {2021})}\BibitemShut {NoStop}%
\bibitem [{\citenamefont {Li}\ \emph {et~al.}(2020)\citenamefont {Li},
  \citenamefont {Li}, \citenamefont {Zhang},\ and\ \citenamefont
  {Gong}}]{li2020symmetry}%
  \BibitemOpen
  \bibfield  {author} {\bibinfo {author} {\bibfnamefont {X.-S.}\ \bibnamefont
  {Li}}, \bibinfo {author} {\bibfnamefont {Z.-Z.}\ \bibnamefont {Li}}, \bibinfo
  {author} {\bibfnamefont {L.-L.}\ \bibnamefont {Zhang}},\ and\ \bibinfo
  {author} {\bibfnamefont {W.-J.}\ \bibnamefont {Gong}},\ }\bibfield  {title}
  {\bibinfo {title} {$\mathcal{PT}$-symmetry of the {S}u-{S}chrieffer-{H}eeger
  model with imaginary boundary potentials and next-nearest-neighboring
  coupling},\ }\href {https://doi.org/10.1088/1361-648x/ab62bd} {\bibfield
  {journal} {\bibinfo  {journal} {Journal of Physics: Condensed Matter}\
  }\textbf {\bibinfo {volume} {32}},\ \bibinfo {pages} {165401} (\bibinfo
  {year} {2020})}\BibitemShut {NoStop}%
\bibitem [{\citenamefont {Yuce}\ and\ \citenamefont
  {Oztas}(2018)}]{yuce2018pt}%
  \BibitemOpen
  \bibfield  {author} {\bibinfo {author} {\bibfnamefont {C.}~\bibnamefont
  {Yuce}}\ and\ \bibinfo {author} {\bibfnamefont {Z.}~\bibnamefont {Oztas}},\
  }\bibfield  {title} {\bibinfo {title} {{PT} symmetry protected
  non-{H}ermitian topological systems},\ }\href
  {https://doi.org/10.1038/s41598-018-35795-5} {\bibfield  {journal} {\bibinfo
  {journal} {Scientific reports}\ }\textbf {\bibinfo {volume} {8}},\ \bibinfo
  {pages} {1} (\bibinfo {year} {2018})}\BibitemShut {NoStop}%
\bibitem [{\citenamefont {Jin}\ \emph {et~al.}(2017)\citenamefont {Jin},
  \citenamefont {Wang},\ and\ \citenamefont {Song}}]{jin2017schrieffer}%
  \BibitemOpen
  \bibfield  {author} {\bibinfo {author} {\bibfnamefont {L.}~\bibnamefont
  {Jin}}, \bibinfo {author} {\bibfnamefont {P.}~\bibnamefont {Wang}},\ and\
  \bibinfo {author} {\bibfnamefont {Z.}~\bibnamefont {Song}},\ }\bibfield
  {title} {\bibinfo {title} {Su-{S}chrieffer-{H}eeger chain with one pair of
  {PT}-symmetric defects},\ }\href {https://doi.org/10.1038/s41598-017-06198-9}
  {\bibfield  {journal} {\bibinfo  {journal} {Scientific Reports}\ }\textbf
  {\bibinfo {volume} {7}},\ \bibinfo {pages} {1} (\bibinfo {year}
  {2017})}\BibitemShut {NoStop}%
\bibitem [{\citenamefont {Zhu}\ \emph {et~al.}(2014)\citenamefont {Zhu},
  \citenamefont {L\"u},\ and\ \citenamefont {Chen}}]{zhu2014pt}%
  \BibitemOpen
  \bibfield  {author} {\bibinfo {author} {\bibfnamefont {B.}~\bibnamefont
  {Zhu}}, \bibinfo {author} {\bibfnamefont {R.}~\bibnamefont {L\"u}},\ and\
  \bibinfo {author} {\bibfnamefont {S.}~\bibnamefont {Chen}},\ }\bibfield
  {title} {\bibinfo {title} {$\mathcal{PT}$ symmetry in the non-hermitian
  {S}u-{S}chrieffer-{H}eeger model with complex boundary potentials},\ }\href
  {https://doi.org/10.1103/PhysRevA.89.062102} {\bibfield  {journal} {\bibinfo
  {journal} {Phys. Rev. A}\ }\textbf {\bibinfo {volume} {89}},\ \bibinfo
  {pages} {062102} (\bibinfo {year} {2014})}\BibitemShut {NoStop}%
\bibitem [{\citenamefont {Xu}\ \emph {et~al.}(2020)\citenamefont {Xu},
  \citenamefont {Zhang}, \citenamefont {Chen}, \citenamefont {Fu},\ and\
  \citenamefont {Zhang}}]{xu2020fate}%
  \BibitemOpen
  \bibfield  {author} {\bibinfo {author} {\bibfnamefont {Z.}~\bibnamefont
  {Xu}}, \bibinfo {author} {\bibfnamefont {R.}~\bibnamefont {Zhang}}, \bibinfo
  {author} {\bibfnamefont {S.}~\bibnamefont {Chen}}, \bibinfo {author}
  {\bibfnamefont {L.}~\bibnamefont {Fu}},\ and\ \bibinfo {author}
  {\bibfnamefont {Y.}~\bibnamefont {Zhang}},\ }\bibfield  {title} {\bibinfo
  {title} {Fate of zero modes in a finite {S}u-{S}chrieffer-{H}eeger model with
  $\mathcal{PT}$ symmetry},\ }\href
  {https://doi.org/10.1103/PhysRevA.101.013635} {\bibfield  {journal} {\bibinfo
   {journal} {Phys. Rev. A}\ }\textbf {\bibinfo {volume} {101}},\ \bibinfo
  {pages} {013635} (\bibinfo {year} {2020})}\BibitemShut {NoStop}%
\bibitem [{\citenamefont {Borgnia}\ \emph {et~al.}(2020)\citenamefont
  {Borgnia}, \citenamefont {Kruchkov},\ and\ \citenamefont
  {Slager}}]{PhysRevLett.124.056802}%
  \BibitemOpen
  \bibfield  {author} {\bibinfo {author} {\bibfnamefont {D.~S.}\ \bibnamefont
  {Borgnia}}, \bibinfo {author} {\bibfnamefont {A.~J.}\ \bibnamefont
  {Kruchkov}},\ and\ \bibinfo {author} {\bibfnamefont {R.-J.}\ \bibnamefont
  {Slager}},\ }\bibfield  {title} {\bibinfo {title} {Non-hermitian boundary
  modes and topology},\ }\href {https://doi.org/10.1103/PhysRevLett.124.056802}
  {\bibfield  {journal} {\bibinfo  {journal} {Phys. Rev. Lett.}\ }\textbf
  {\bibinfo {volume} {124}},\ \bibinfo {pages} {056802} (\bibinfo {year}
  {2020})}\BibitemShut {NoStop}%
\bibitem [{\citenamefont {Okuma}\ \emph {et~al.}(2020)\citenamefont {Okuma},
  \citenamefont {Kawabata}, \citenamefont {Shiozaki},\ and\ \citenamefont
  {Sato}}]{PhysRevLett.124.086801}%
  \BibitemOpen
  \bibfield  {author} {\bibinfo {author} {\bibfnamefont {N.}~\bibnamefont
  {Okuma}}, \bibinfo {author} {\bibfnamefont {K.}~\bibnamefont {Kawabata}},
  \bibinfo {author} {\bibfnamefont {K.}~\bibnamefont {Shiozaki}},\ and\
  \bibinfo {author} {\bibfnamefont {M.}~\bibnamefont {Sato}},\ }\bibfield
  {title} {\bibinfo {title} {Topological origin of non-hermitian skin
  effects},\ }\href {https://doi.org/10.1103/PhysRevLett.124.086801} {\bibfield
   {journal} {\bibinfo  {journal} {Phys. Rev. Lett.}\ }\textbf {\bibinfo
  {volume} {124}},\ \bibinfo {pages} {086801} (\bibinfo {year}
  {2020})}\BibitemShut {NoStop}%
\bibitem [{\citenamefont {Zhang}\ \emph {et~al.}(2020)\citenamefont {Zhang},
  \citenamefont {Yang},\ and\ \citenamefont {Fang}}]{PhysRevLett.125.126402}%
  \BibitemOpen
  \bibfield  {author} {\bibinfo {author} {\bibfnamefont {K.}~\bibnamefont
  {Zhang}}, \bibinfo {author} {\bibfnamefont {Z.}~\bibnamefont {Yang}},\ and\
  \bibinfo {author} {\bibfnamefont {C.}~\bibnamefont {Fang}},\ }\bibfield
  {title} {\bibinfo {title} {Correspondence between winding numbers and skin
  modes in non-hermitian systems},\ }\href
  {https://doi.org/10.1103/PhysRevLett.125.126402} {\bibfield  {journal}
  {\bibinfo  {journal} {Phys. Rev. Lett.}\ }\textbf {\bibinfo {volume} {125}},\
  \bibinfo {pages} {126402} (\bibinfo {year} {2020})}\BibitemShut {NoStop}%
\bibitem [{\citenamefont {Yao}\ and\ \citenamefont {Wang}(2018)}]{yao2018edge}%
  \BibitemOpen
  \bibfield  {author} {\bibinfo {author} {\bibfnamefont {S.}~\bibnamefont
  {Yao}}\ and\ \bibinfo {author} {\bibfnamefont {Z.}~\bibnamefont {Wang}},\
  }\bibfield  {title} {\bibinfo {title} {Edge states and topological invariants
  of non-{H}ermitian systems},\ }\href
  {https://doi.org/10.1103/PhysRevLett.121.086803} {\bibfield  {journal}
  {\bibinfo  {journal} {Phys. Rev. Lett.}\ }\textbf {\bibinfo {volume} {121}},\
  \bibinfo {pages} {086803} (\bibinfo {year} {2018})}\BibitemShut {NoStop}%
\bibitem [{\citenamefont {Song}\ \emph {et~al.}(2019)\citenamefont {Song},
  \citenamefont {Yao},\ and\ \citenamefont {Wang}}]{song2019non}%
  \BibitemOpen
  \bibfield  {author} {\bibinfo {author} {\bibfnamefont {F.}~\bibnamefont
  {Song}}, \bibinfo {author} {\bibfnamefont {S.}~\bibnamefont {Yao}},\ and\
  \bibinfo {author} {\bibfnamefont {Z.}~\bibnamefont {Wang}},\ }\bibfield
  {title} {\bibinfo {title} {Non-{H}ermitian topological invariants in real
  space},\ }\href {https://doi.org/10.1103/PhysRevLett.123.246801} {\bibfield
  {journal} {\bibinfo  {journal} {Phys. Rev. Lett.}\ }\textbf {\bibinfo
  {volume} {123}},\ \bibinfo {pages} {246801} (\bibinfo {year}
  {2019})}\BibitemShut {NoStop}%
\bibitem [{\citenamefont {Lee}(2016)}]{lee2016anomalous}%
  \BibitemOpen
  \bibfield  {author} {\bibinfo {author} {\bibfnamefont {T.~E.}\ \bibnamefont
  {Lee}},\ }\bibfield  {title} {\bibinfo {title} {Anomalous edge state in a
  non-{H}ermitian lattice},\ }\href
  {https://doi.org/10.1103/PhysRevLett.116.133903} {\bibfield  {journal}
  {\bibinfo  {journal} {Phys. Rev. Lett.}\ }\textbf {\bibinfo {volume} {116}},\
  \bibinfo {pages} {133903} (\bibinfo {year} {2016})}\BibitemShut {NoStop}%
\bibitem [{\citenamefont {Liang}\ \emph {et~al.}(2014)\citenamefont {Liang},
  \citenamefont {Scott},\ and\ \citenamefont {Joglekar}}]{PhysRevA.89.030102}%
  \BibitemOpen
  \bibfield  {author} {\bibinfo {author} {\bibfnamefont {C.~H.}\ \bibnamefont
  {Liang}}, \bibinfo {author} {\bibfnamefont {D.~D.}\ \bibnamefont {Scott}},\
  and\ \bibinfo {author} {\bibfnamefont {Y.~N.}\ \bibnamefont {Joglekar}},\
  }\bibfield  {title} {\bibinfo {title} {$\mathcal{PT}$ restoration via
  increased loss and gain in the $\mathcal{PT}$-symmetric aubry-andr\'e
  model},\ }\href {https://doi.org/10.1103/PhysRevA.89.030102} {\bibfield
  {journal} {\bibinfo  {journal} {Phys. Rev. A}\ }\textbf {\bibinfo {volume}
  {89}},\ \bibinfo {pages} {030102} (\bibinfo {year} {2014})}\BibitemShut
  {NoStop}%
\bibitem [{\citenamefont {Yuce}(2014)}]{YUCE20142024}%
  \BibitemOpen
  \bibfield  {author} {\bibinfo {author} {\bibfnamefont {C.}~\bibnamefont
  {Yuce}},\ }\bibfield  {title} {\bibinfo {title} {Pt symmetric aubry–andre
  model},\ }\href
  {https://doi.org/https://doi.org/10.1016/j.physleta.2014.05.005} {\bibfield
  {journal} {\bibinfo  {journal} {Physics Letters A}\ }\textbf {\bibinfo
  {volume} {378}},\ \bibinfo {pages} {2024} (\bibinfo {year}
  {2014})}\BibitemShut {NoStop}%
\bibitem [{\citenamefont {Schiffer}\ \emph {et~al.}(2021)\citenamefont
  {Schiffer}, \citenamefont {Liu}, \citenamefont {Hu},\ and\ \citenamefont
  {Wang}}]{PhysRevA.103.L011302}%
  \BibitemOpen
  \bibfield  {author} {\bibinfo {author} {\bibfnamefont {S.}~\bibnamefont
  {Schiffer}}, \bibinfo {author} {\bibfnamefont {X.-J.}\ \bibnamefont {Liu}},
  \bibinfo {author} {\bibfnamefont {H.}~\bibnamefont {Hu}},\ and\ \bibinfo
  {author} {\bibfnamefont {J.}~\bibnamefont {Wang}},\ }\bibfield  {title}
  {\bibinfo {title} {Anderson localization transition in a robust
  $\mathcal{PT}$-symmetric phase of a generalized aubry-andr\'e model},\ }\href
  {https://doi.org/10.1103/PhysRevA.103.L011302} {\bibfield  {journal}
  {\bibinfo  {journal} {Phys. Rev. A}\ }\textbf {\bibinfo {volume} {103}},\
  \bibinfo {pages} {L011302} (\bibinfo {year} {2021})}\BibitemShut {NoStop}%
\bibitem [{\citenamefont {Yuce}(2015)}]{YUCE20151213}%
  \BibitemOpen
  \bibfield  {author} {\bibinfo {author} {\bibfnamefont {C.}~\bibnamefont
  {Yuce}},\ }\bibfield  {title} {\bibinfo {title} {Topological phase in a
  non-hermitian pt symmetric system},\ }\href
  {https://doi.org/https://doi.org/10.1016/j.physleta.2015.02.011} {\bibfield
  {journal} {\bibinfo  {journal} {Physics Letters A}\ }\textbf {\bibinfo
  {volume} {379}},\ \bibinfo {pages} {1213} (\bibinfo {year}
  {2015})}\BibitemShut {NoStop}%
\bibitem [{\citenamefont {Harter}\ \emph {et~al.}(2016)\citenamefont {Harter},
  \citenamefont {Lee},\ and\ \citenamefont {Joglekar}}]{PhysRevA.93.062101}%
  \BibitemOpen
  \bibfield  {author} {\bibinfo {author} {\bibfnamefont {A.~K.}\ \bibnamefont
  {Harter}}, \bibinfo {author} {\bibfnamefont {T.~E.}\ \bibnamefont {Lee}},\
  and\ \bibinfo {author} {\bibfnamefont {Y.~N.}\ \bibnamefont {Joglekar}},\
  }\bibfield  {title} {\bibinfo {title} {$\mathcal{PT}$-breaking threshold in
  spatially asymmetric aubry-andr\'e and harper models: Hidden symmetry and
  topological states},\ }\href {https://doi.org/10.1103/PhysRevA.93.062101}
  {\bibfield  {journal} {\bibinfo  {journal} {Phys. Rev. A}\ }\textbf {\bibinfo
  {volume} {93}},\ \bibinfo {pages} {062101} (\bibinfo {year}
  {2016})}\BibitemShut {NoStop}%
\bibitem [{\citenamefont {Longhi}(2019{\natexlab{a}})}]{longhi2019metal}%
  \BibitemOpen
  \bibfield  {author} {\bibinfo {author} {\bibfnamefont {S.}~\bibnamefont
  {Longhi}},\ }\bibfield  {title} {\bibinfo {title} {Metal-insulator phase
  transition in a non-{H}ermitian {A}ubry-{A}ndr\'e-{H}arper model},\ }\href
  {https://doi.org/10.1103/PhysRevB.100.125157} {\bibfield  {journal} {\bibinfo
   {journal} {Phys. Rev. B}\ }\textbf {\bibinfo {volume} {100}},\ \bibinfo
  {pages} {125157} (\bibinfo {year} {2019}{\natexlab{a}})}\BibitemShut
  {NoStop}%
\bibitem [{\citenamefont {Longhi}(2019{\natexlab{b}})}]{longhi2019topological}%
  \BibitemOpen
  \bibfield  {author} {\bibinfo {author} {\bibfnamefont {S.}~\bibnamefont
  {Longhi}},\ }\bibfield  {title} {\bibinfo {title} {Topological phase
  transition in non-{H}ermitian quasicrystals},\ }\href
  {https://doi.org/10.1103/PhysRevLett.122.237601} {\bibfield  {journal}
  {\bibinfo  {journal} {Phys. Rev. Lett.}\ }\textbf {\bibinfo {volume} {122}},\
  \bibinfo {pages} {237601} (\bibinfo {year} {2019}{\natexlab{b}})}\BibitemShut
  {NoStop}%
\bibitem [{\citenamefont {Liu}\ \emph {et~al.}(2021{\natexlab{a}})\citenamefont
  {Liu}, \citenamefont {Zhou},\ and\ \citenamefont {Chen}}]{Shu-Chen}%
  \BibitemOpen
  \bibfield  {author} {\bibinfo {author} {\bibfnamefont {Y.}~\bibnamefont
  {Liu}}, \bibinfo {author} {\bibfnamefont {Q.}~\bibnamefont {Zhou}},\ and\
  \bibinfo {author} {\bibfnamefont {S.}~\bibnamefont {Chen}},\ }\bibfield
  {title} {\bibinfo {title} {Localization transition, spectrum structure, and
  winding numbers for one-dimensional non-hermitian quasicrystals},\ }\href
  {https://doi.org/10.1103/PhysRevB.104.024201} {\bibfield  {journal} {\bibinfo
   {journal} {Phys. Rev. B}\ }\textbf {\bibinfo {volume} {104}},\ \bibinfo
  {pages} {024201} (\bibinfo {year} {2021}{\natexlab{a}})}\BibitemShut
  {NoStop}%
\bibitem [{\citenamefont {Tang}\ \emph {et~al.}(2021)\citenamefont {Tang},
  \citenamefont {Zhang}, \citenamefont {Zhang},\ and\ \citenamefont
  {Zhang}}]{PhysRevA.103.033325}%
  \BibitemOpen
  \bibfield  {author} {\bibinfo {author} {\bibfnamefont {L.-Z.}\ \bibnamefont
  {Tang}}, \bibinfo {author} {\bibfnamefont {G.-Q.}\ \bibnamefont {Zhang}},
  \bibinfo {author} {\bibfnamefont {L.-F.}\ \bibnamefont {Zhang}},\ and\
  \bibinfo {author} {\bibfnamefont {D.-W.}\ \bibnamefont {Zhang}},\ }\bibfield
  {title} {\bibinfo {title} {Localization and topological transitions in
  non-hermitian quasiperiodic lattices},\ }\href
  {https://doi.org/10.1103/PhysRevA.103.033325} {\bibfield  {journal} {\bibinfo
   {journal} {Phys. Rev. A}\ }\textbf {\bibinfo {volume} {103}},\ \bibinfo
  {pages} {033325} (\bibinfo {year} {2021})}\BibitemShut {NoStop}%
\bibitem [{\citenamefont {Gandhi}\ and\ \citenamefont
  {Bandyopadhyay}(2023)}]{PhysRevB.108.014204}%
  \BibitemOpen
  \bibfield  {author} {\bibinfo {author} {\bibfnamefont {S.}~\bibnamefont
  {Gandhi}}\ and\ \bibinfo {author} {\bibfnamefont {J.~N.}\ \bibnamefont
  {Bandyopadhyay}},\ }\bibfield  {title} {\bibinfo {title} {Topological triple
  phase transition in non-hermitian quasicrystals with complex asymmetric
  hopping},\ }\href {https://doi.org/10.1103/PhysRevB.108.014204} {\bibfield
  {journal} {\bibinfo  {journal} {Phys. Rev. B}\ }\textbf {\bibinfo {volume}
  {108}},\ \bibinfo {pages} {014204} (\bibinfo {year} {2023})}\BibitemShut
  {NoStop}%
\bibitem [{\citenamefont {Padhan}\ \emph {et~al.}(2024)\citenamefont {Padhan},
  \citenamefont {Padhi},\ and\ \citenamefont {Mishra}}]{PhysRevB.109.L020203}%
  \BibitemOpen
  \bibfield  {author} {\bibinfo {author} {\bibfnamefont {A.}~\bibnamefont
  {Padhan}}, \bibinfo {author} {\bibfnamefont {S.~R.}\ \bibnamefont {Padhi}},\
  and\ \bibinfo {author} {\bibfnamefont {T.}~\bibnamefont {Mishra}},\
  }\bibfield  {title} {\bibinfo {title} {Complete delocalization and reentrant
  topological transition in a non-hermitian quasiperiodic lattice},\ }\href
  {https://doi.org/10.1103/PhysRevB.109.L020203} {\bibfield  {journal}
  {\bibinfo  {journal} {Phys. Rev. B}\ }\textbf {\bibinfo {volume} {109}},\
  \bibinfo {pages} {L020203} (\bibinfo {year} {2024})}\BibitemShut {NoStop}%
\bibitem [{\citenamefont {Zeng}\ and\ \citenamefont
  {Xu}(2020)}]{PhysRevResearch.2.033052}%
  \BibitemOpen
  \bibfield  {author} {\bibinfo {author} {\bibfnamefont {Q.-B.}\ \bibnamefont
  {Zeng}}\ and\ \bibinfo {author} {\bibfnamefont {Y.}~\bibnamefont {Xu}},\
  }\bibfield  {title} {\bibinfo {title} {Winding numbers and generalized
  mobility edges in non-hermitian systems},\ }\href
  {https://doi.org/10.1103/PhysRevResearch.2.033052} {\bibfield  {journal}
  {\bibinfo  {journal} {Phys. Rev. Res.}\ }\textbf {\bibinfo {volume} {2}},\
  \bibinfo {pages} {033052} (\bibinfo {year} {2020})}\BibitemShut {NoStop}%
\bibitem [{\citenamefont {Zeng}\ \emph {et~al.}(2020)\citenamefont {Zeng},
  \citenamefont {Yang},\ and\ \citenamefont {Xu}}]{PhysRevB.101.020201}%
  \BibitemOpen
  \bibfield  {author} {\bibinfo {author} {\bibfnamefont {Q.-B.}\ \bibnamefont
  {Zeng}}, \bibinfo {author} {\bibfnamefont {Y.-B.}\ \bibnamefont {Yang}},\
  and\ \bibinfo {author} {\bibfnamefont {Y.}~\bibnamefont {Xu}},\ }\bibfield
  {title} {\bibinfo {title} {Topological phases in non-hermitian
  aubry-andr\'e-harper models},\ }\href
  {https://doi.org/10.1103/PhysRevB.101.020201} {\bibfield  {journal} {\bibinfo
   {journal} {Phys. Rev. B}\ }\textbf {\bibinfo {volume} {101}},\ \bibinfo
  {pages} {020201} (\bibinfo {year} {2020})}\BibitemShut {NoStop}%
\bibitem [{\citenamefont {Wang}\ \emph {et~al.}(2022)\citenamefont {Wang},
  \citenamefont {Xu}, \citenamefont {Li}, \citenamefont {Xu},\ and\
  \citenamefont {Wang}}]{PhysRevB.105.024514}%
  \BibitemOpen
  \bibfield  {author} {\bibinfo {author} {\bibfnamefont {Z.-H.}\ \bibnamefont
  {Wang}}, \bibinfo {author} {\bibfnamefont {F.}~\bibnamefont {Xu}}, \bibinfo
  {author} {\bibfnamefont {L.}~\bibnamefont {Li}}, \bibinfo {author}
  {\bibfnamefont {D.-H.}\ \bibnamefont {Xu}},\ and\ \bibinfo {author}
  {\bibfnamefont {B.}~\bibnamefont {Wang}},\ }\bibfield  {title} {\bibinfo
  {title} {Topological superconductors and exact mobility edges in
  non-hermitian quasicrystals},\ }\href
  {https://doi.org/10.1103/PhysRevB.105.024514} {\bibfield  {journal} {\bibinfo
   {journal} {Phys. Rev. B}\ }\textbf {\bibinfo {volume} {105}},\ \bibinfo
  {pages} {024514} (\bibinfo {year} {2022})}\BibitemShut {NoStop}%
\bibitem [{\citenamefont {Liu}\ \emph {et~al.}(2021{\natexlab{b}})\citenamefont
  {Liu}, \citenamefont {Cheng}, \citenamefont {Guo},\ and\ \citenamefont
  {Xianlong}}]{PhysRevB.103.104203}%
  \BibitemOpen
  \bibfield  {author} {\bibinfo {author} {\bibfnamefont {T.}~\bibnamefont
  {Liu}}, \bibinfo {author} {\bibfnamefont {S.}~\bibnamefont {Cheng}}, \bibinfo
  {author} {\bibfnamefont {H.}~\bibnamefont {Guo}},\ and\ \bibinfo {author}
  {\bibfnamefont {G.}~\bibnamefont {Xianlong}},\ }\bibfield  {title} {\bibinfo
  {title} {Fate of majorana zero modes, exact location of critical states, and
  unconventional real-complex transition in non-hermitian quasiperiodic
  lattices},\ }\href {https://doi.org/10.1103/PhysRevB.103.104203} {\bibfield
  {journal} {\bibinfo  {journal} {Phys. Rev. B}\ }\textbf {\bibinfo {volume}
  {103}},\ \bibinfo {pages} {104203} (\bibinfo {year}
  {2021}{\natexlab{b}})}\BibitemShut {NoStop}%
\bibitem [{\citenamefont {Cai}(2021)}]{PhysRevB.103.214202}%
  \BibitemOpen
  \bibfield  {author} {\bibinfo {author} {\bibfnamefont {X.}~\bibnamefont
  {Cai}},\ }\bibfield  {title} {\bibinfo {title} {Localization and topological
  phase transitions in non-hermitian aubry-andr\'e-harper models with $p$-wave
  pairing},\ }\href {https://doi.org/10.1103/PhysRevB.103.214202} {\bibfield
  {journal} {\bibinfo  {journal} {Phys. Rev. B}\ }\textbf {\bibinfo {volume}
  {103}},\ \bibinfo {pages} {214202} (\bibinfo {year} {2021})}\BibitemShut
  {NoStop}%
\bibitem [{\citenamefont {Cheng}\ and\ \citenamefont {Gao}(2022)}]{Cheng_Gao}%
  \BibitemOpen
  \bibfield  {author} {\bibinfo {author} {\bibfnamefont {S.}~\bibnamefont
  {Cheng}}\ and\ \bibinfo {author} {\bibfnamefont {X.}~\bibnamefont {Gao}},\
  }\bibfield  {title} {\bibinfo {title} {Majorana zero modes, unconventional
  real-complex transition, and mobility edges in a one-dimensional
  non-hermitian quasi-periodic lattice},\ }\href
  {https://doi.org/10.1088/1674-1056/ac3222} {\bibfield  {journal} {\bibinfo
  {journal} {Chinese Physics B}\ }\textbf {\bibinfo {volume} {31}},\ \bibinfo
  {pages} {017401} (\bibinfo {year} {2022})},\ \Eprint
  {https://arxiv.org/abs/2105.06621} {arXiv:2105.06621 [cond-mat.dis-nn]}
  \BibitemShut {NoStop}%
\bibitem [{\citenamefont {Jiang}\ \emph {et~al.}(2021)\citenamefont {Jiang},
  \citenamefont {Qiao},\ and\ \citenamefont {Cao}}]{Xiang_Ping_Jiang}%
  \BibitemOpen
  \bibfield  {author} {\bibinfo {author} {\bibfnamefont {X.-P.}\ \bibnamefont
  {Jiang}}, \bibinfo {author} {\bibfnamefont {Y.}~\bibnamefont {Qiao}},\ and\
  \bibinfo {author} {\bibfnamefont {J.}~\bibnamefont {Cao}},\ }\bibfield
  {title} {\bibinfo {title} {Non-hermitian kitaev chain with complex periodic
  and quasiperiodic potentials},\ }\href
  {https://doi.org/10.1088/1674-1056/abfa08} {\bibfield  {journal} {\bibinfo
  {journal} {Chinese Physics B}\ }\textbf {\bibinfo {volume} {30}},\ \bibinfo
  {pages} {077101} (\bibinfo {year} {2021})}\BibitemShut {NoStop}%
\bibitem [{\citenamefont {Bender}\ and\ \citenamefont
  {Boettcher}(1998)}]{bender1998real}%
  \BibitemOpen
  \bibfield  {author} {\bibinfo {author} {\bibfnamefont {C.~M.}\ \bibnamefont
  {Bender}}\ and\ \bibinfo {author} {\bibfnamefont {S.}~\bibnamefont
  {Boettcher}},\ }\bibfield  {title} {\bibinfo {title} {Real spectra in
  non-{H}ermitian {H}amiltonians having $\mathcal{P}\mathcal{T}$ symmetry},\
  }\href {https://doi.org/10.1103/PhysRevLett.80.5243} {\bibfield  {journal}
  {\bibinfo  {journal} {Phys. Rev. Lett.}\ }\textbf {\bibinfo {volume} {80}},\
  \bibinfo {pages} {5243} (\bibinfo {year} {1998})}\BibitemShut {NoStop}%
\bibitem [{\citenamefont {El-Ganainy}\ \emph {et~al.}(2018)\citenamefont
  {El-Ganainy}, \citenamefont {Makris}, \citenamefont {Khajavikhan},
  \citenamefont {Musslimani}, \citenamefont {Rotter},\ and\ \citenamefont
  {Christodoulides}}]{el2018non}%
  \BibitemOpen
  \bibfield  {author} {\bibinfo {author} {\bibfnamefont {R.}~\bibnamefont
  {El-Ganainy}}, \bibinfo {author} {\bibfnamefont {K.~G.}\ \bibnamefont
  {Makris}}, \bibinfo {author} {\bibfnamefont {M.}~\bibnamefont {Khajavikhan}},
  \bibinfo {author} {\bibfnamefont {Z.~H.}\ \bibnamefont {Musslimani}},
  \bibinfo {author} {\bibfnamefont {S.}~\bibnamefont {Rotter}},\ and\ \bibinfo
  {author} {\bibfnamefont {D.~N.}\ \bibnamefont {Christodoulides}},\ }\bibfield
   {title} {\bibinfo {title} {Non-{H}ermitian physics and {PT} symmetry},\
  }\href {https://doi.org/10.1038/nphys4323} {\bibfield  {journal} {\bibinfo
  {journal} {Nature Physics}\ }\textbf {\bibinfo {volume} {14}},\ \bibinfo
  {pages} {11} (\bibinfo {year} {2018})}\BibitemShut {NoStop}%
\bibitem [{\citenamefont {Bender}\ \emph {et~al.}(2002)\citenamefont {Bender},
  \citenamefont {Berry},\ and\ \citenamefont
  {Mandilara}}]{bender2002generalized}%
  \BibitemOpen
  \bibfield  {author} {\bibinfo {author} {\bibfnamefont {C.~M.}\ \bibnamefont
  {Bender}}, \bibinfo {author} {\bibfnamefont {M.~V.}\ \bibnamefont {Berry}},\
  and\ \bibinfo {author} {\bibfnamefont {A.}~\bibnamefont {Mandilara}},\
  }\bibfield  {title} {\bibinfo {title} {Generalized {PT} symmetry and real
  spectra},\ }\href {https://doi.org/10.1088/0305-4470/35/31/101} {\bibfield
  {journal} {\bibinfo  {journal} {Journal of Physics A: Mathematical and
  General}\ }\textbf {\bibinfo {volume} {35}},\ \bibinfo {pages} {L467}
  (\bibinfo {year} {2002})}\BibitemShut {NoStop}%
\bibitem [{\citenamefont {Hegde}\ and\ \citenamefont
  {Vishveshwara}(2016)}]{PhysRevB.94.115166}%
  \BibitemOpen
  \bibfield  {author} {\bibinfo {author} {\bibfnamefont {S.~S.}\ \bibnamefont
  {Hegde}}\ and\ \bibinfo {author} {\bibfnamefont {S.}~\bibnamefont
  {Vishveshwara}},\ }\bibfield  {title} {\bibinfo {title} {Majorana
  wave-function oscillations, fermion parity switches, and disorder in kitaev
  chains},\ }\href {https://doi.org/10.1103/PhysRevB.94.115166} {\bibfield
  {journal} {\bibinfo  {journal} {Phys. Rev. B}\ }\textbf {\bibinfo {volume}
  {94}},\ \bibinfo {pages} {115166} (\bibinfo {year} {2016})}\BibitemShut
  {NoStop}%
\bibitem [{\citenamefont {Wimmer}(2012)}]{Wimmer}%
  \BibitemOpen
  \bibfield  {author} {\bibinfo {author} {\bibfnamefont {M.}~\bibnamefont
  {Wimmer}},\ }\bibfield  {title} {\bibinfo {title} {Algorithm 923: Efficient
  numerical computation of the pfaffian for dense and banded skew-symmetric
  matrices},\ }\bibfield  {journal} {\bibinfo  {journal} {ACM Trans. Math.
  Softw.}\ }\textbf {\bibinfo {volume} {38}},\ \href
  {https://doi.org/10.1145/2331130.2331138} {10.1145/2331130.2331138} (\bibinfo
  {year} {2012})\BibitemShut {NoStop}%
\bibitem [{\citenamefont {Kawabata}\ \emph {et~al.}(2019)\citenamefont
  {Kawabata}, \citenamefont {Shiozaki}, \citenamefont {Ueda},\ and\
  \citenamefont {Sato}}]{PhysRevX.9.041015}%
  \BibitemOpen
  \bibfield  {author} {\bibinfo {author} {\bibfnamefont {K.}~\bibnamefont
  {Kawabata}}, \bibinfo {author} {\bibfnamefont {K.}~\bibnamefont {Shiozaki}},
  \bibinfo {author} {\bibfnamefont {M.}~\bibnamefont {Ueda}},\ and\ \bibinfo
  {author} {\bibfnamefont {M.}~\bibnamefont {Sato}},\ }\bibfield  {title}
  {\bibinfo {title} {Symmetry and topology in non-hermitian physics},\ }\href
  {https://doi.org/10.1103/PhysRevX.9.041015} {\bibfield  {journal} {\bibinfo
  {journal} {Phys. Rev. X}\ }\textbf {\bibinfo {volume} {9}},\ \bibinfo {pages}
  {041015} (\bibinfo {year} {2019})}\BibitemShut {NoStop}%
\bibitem [{\citenamefont {Schreiber}\ and\ \citenamefont
  {Grussbach}(1991)}]{PhysRevLett.67.607}%
  \BibitemOpen
  \bibfield  {author} {\bibinfo {author} {\bibfnamefont {M.}~\bibnamefont
  {Schreiber}}\ and\ \bibinfo {author} {\bibfnamefont {H.}~\bibnamefont
  {Grussbach}},\ }\bibfield  {title} {\bibinfo {title} {Multifractal wave
  functions at the anderson transition},\ }\href
  {https://doi.org/10.1103/PhysRevLett.67.607} {\bibfield  {journal} {\bibinfo
  {journal} {Phys. Rev. Lett.}\ }\textbf {\bibinfo {volume} {67}},\ \bibinfo
  {pages} {607} (\bibinfo {year} {1991})}\BibitemShut {NoStop}%
\bibitem [{\citenamefont {Fraxanet}\ \emph {et~al.}(2022)\citenamefont
  {Fraxanet}, \citenamefont {Bhattacharya}, \citenamefont {Grass},
  \citenamefont {Lewenstein},\ and\ \citenamefont
  {Dauphin}}]{PhysRevB.106.024204}%
  \BibitemOpen
  \bibfield  {author} {\bibinfo {author} {\bibfnamefont {J.}~\bibnamefont
  {Fraxanet}}, \bibinfo {author} {\bibfnamefont {U.}~\bibnamefont
  {Bhattacharya}}, \bibinfo {author} {\bibfnamefont {T.}~\bibnamefont {Grass}},
  \bibinfo {author} {\bibfnamefont {M.}~\bibnamefont {Lewenstein}},\ and\
  \bibinfo {author} {\bibfnamefont {A.}~\bibnamefont {Dauphin}},\ }\bibfield
  {title} {\bibinfo {title} {Localization and multifractal properties of the
  long-range kitaev chain in the presence of an aubry-andr\'e-harper
  modulation},\ }\href {https://doi.org/10.1103/PhysRevB.106.024204} {\bibfield
   {journal} {\bibinfo  {journal} {Phys. Rev. B}\ }\textbf {\bibinfo {volume}
  {106}},\ \bibinfo {pages} {024204} (\bibinfo {year} {2022})}\BibitemShut
  {NoStop}%
\bibitem [{\citenamefont {Deng}\ \emph {et~al.}(2019)\citenamefont {Deng},
  \citenamefont {Ray}, \citenamefont {Sinha}, \citenamefont {Shlyapnikov},\
  and\ \citenamefont {Santos}}]{PhysRevLett.123.025301}%
  \BibitemOpen
  \bibfield  {author} {\bibinfo {author} {\bibfnamefont {X.}~\bibnamefont
  {Deng}}, \bibinfo {author} {\bibfnamefont {S.}~\bibnamefont {Ray}}, \bibinfo
  {author} {\bibfnamefont {S.}~\bibnamefont {Sinha}}, \bibinfo {author}
  {\bibfnamefont {G.~V.}\ \bibnamefont {Shlyapnikov}},\ and\ \bibinfo {author}
  {\bibfnamefont {L.}~\bibnamefont {Santos}},\ }\bibfield  {title} {\bibinfo
  {title} {One-dimensional quasicrystals with power-law hopping},\ }\href
  {https://doi.org/10.1103/PhysRevLett.123.025301} {\bibfield  {journal}
  {\bibinfo  {journal} {Phys. Rev. Lett.}\ }\textbf {\bibinfo {volume} {123}},\
  \bibinfo {pages} {025301} (\bibinfo {year} {2019})}\BibitemShut {NoStop}%
\bibitem [{\citenamefont {Roy}\ and\ \citenamefont
  {Sharma}(2021)}]{PhysRevB.103.075124}%
  \BibitemOpen
  \bibfield  {author} {\bibinfo {author} {\bibfnamefont {N.}~\bibnamefont
  {Roy}}\ and\ \bibinfo {author} {\bibfnamefont {A.}~\bibnamefont {Sharma}},\
  }\bibfield  {title} {\bibinfo {title} {Fraction of delocalized eigenstates in
  the long-range aubry-andr\'e-harper model},\ }\href
  {https://doi.org/10.1103/PhysRevB.103.075124} {\bibfield  {journal} {\bibinfo
   {journal} {Phys. Rev. B}\ }\textbf {\bibinfo {volume} {103}},\ \bibinfo
  {pages} {075124} (\bibinfo {year} {2021})}\BibitemShut {NoStop}%
\bibitem [{\citenamefont {Peng}\ \emph {et~al.}(2023)\citenamefont {Peng},
  \citenamefont {Cheng},\ and\ \citenamefont {Xianlong}}]{PhysRevB.107.174205}%
  \BibitemOpen
  \bibfield  {author} {\bibinfo {author} {\bibfnamefont {D.}~\bibnamefont
  {Peng}}, \bibinfo {author} {\bibfnamefont {S.}~\bibnamefont {Cheng}},\ and\
  \bibinfo {author} {\bibfnamefont {G.}~\bibnamefont {Xianlong}},\ }\bibfield
  {title} {\bibinfo {title} {Power law hopping of single particles in
  one-dimensional non-hermitian quasicrystals},\ }\href
  {https://doi.org/10.1103/PhysRevB.107.174205} {\bibfield  {journal} {\bibinfo
   {journal} {Phys. Rev. B}\ }\textbf {\bibinfo {volume} {107}},\ \bibinfo
  {pages} {174205} (\bibinfo {year} {2023})}\BibitemShut {NoStop}%
\end{thebibliography}
\end{document}